\documentclass[american,aps,twocolumn,showkeys,superscriptaddress,prr,longbibliography]{revtex4-2}

\usepackage[T1]{fontenc}

\usepackage[latin9]{inputenc}
\usepackage{geometry}
\usepackage{color}
\definecolor{page_backgroundcolor}{rgb}{1, 1, 1}
\pagecolor{page_backgroundcolor}
\definecolor{document_fontcolor}{rgb}{0, 0, 0}
\color{document_fontcolor}
\usepackage{float}
\usepackage{dsfont}
\usepackage{amsmath}
\usepackage{amssymb}
\usepackage{graphicx}
\usepackage{wasysym}
\usepackage{microtype}
\usepackage{pstricks}
\usepackage{multirow}

\makeatletter

\providecommand{\tabularnewline}{\\}
\usepackage[low-sup]{subdepth}

\usepackage[colorlinks=true,linkcolor=red,citecolor=blue,urlcolor=blue]{hyperref}
\usepackage{orcidlink}

\AtBeginDocument{
  
}

\makeatother

\usepackage{babel}
\begin{document}
\title{Phase-Driven Precision Boost in Quantum Compression for Postselected
Metrology}
\author{Aiham M. Rostom~\!\!\orcidlink{0000-0002-0084-0961}}
\email{a.rostom@g.nsu.ru}
\affiliation{Novosibirsk State University, 630090, Novosibirsk, Russia}
\affiliation{Institute of Automation and Electrometry SBRAS, 630090, Novosibirsk,
Russia}
\affiliation{Department of Physics, Lattakia University, Lattakia, Syria}

\author{Saeed Haddadi~\!\!\orcidlink{0000-0002-1596-0763}}
\email{haddadi@ipm.ir}
\affiliation{School of Particles and Accelerators, Institute for Research in Fundamental
Sciences (IPM), P.O. Box 19395-5531, Tehran, Iran}

\author{Vladimir A. Tomilin~\!\!\orcidlink{ }}
\affiliation{Institute of Automation and Electrometry SBRAS, 630090, Novosibirsk,
Russia}


\begin{abstract}
We reveal the noncyclic Pancharatnam phase--arising from the coherent system-meter interaction--as a fundamental criterion that governs the optimal performance of quantum compression channels in postselected metrology. This phase embodies a phase connection that enables precise control over the parallel evolution of the meter state, thereby maximizing the quantum Fisher information per trial and achieving lossless compression channels. Remarkably, fine-tuning the postselection parameter
just below this optimal phase incurs substantial information loss,
whereas tuning it just above fully suppresses undesired parallel evolution,  enhancing information retention beyond that achievable in postselected protocols lacking Pancharatnam phase effects. We further reveal that leveraging qudit meter states can unlock a substantial additional enhancement. These
findings establish the Pancharatnam phase as a
geometric benchmark, guiding the design of high-precision
quantum parameter estimation protocols.
\end{abstract}
\maketitle

\section{Introduction}
Harnessing quantum correlations, quantum metrology enables
parameter estimation at sensitivities unattainable by classical methods,
with broad implications for foundational physics and advanced measurement
applications \citep{Taylor2016,Giovannetti2011Advances,Pezze2018,DeMille2024,Wu2024,Yurischev2025}.
Compression channels in postselected quantum metrology, wherein measurement outcomes are conditioned on specific postselection events, have emerged as a powerful paradigm for amplifying weak signals and enhancing sensitivity, offering genuine  quantum and technical advantages \citep{ArvidssonShukur2020,Salvati2024,Yang2024}.

Quantum compression channels enable more efficient and
robust quantum metrology by allowing information to be concentrated
into fewer, higher-quality measurement outcomes that are particularly valuable
when measurements are expensive or noisy \citep{ArvidssonShukur2020,Hallaji2017,Arvidsson-Shukur_2024}.

By employing postselection protocols, compression channels
offer a range of significant technical advantages for enhancing precision
of quantum metrology. Among these is the effective amplification
of the input quanta flux in systems characterized by low generation
rates, achieved through the recycling of unpostselected quantum resources
\citep{Dressel2013}. Postselection inherently circumvents detector
saturation by selectively heralding rare  events \citep{Harris2017},
thereby mitigating nonlinear detector response and dead-time limitations.
Moreover, the exploitation of postselection frameworks enhances the
discrimination fidelity between genuine single photons, substantially
suppressing detector noise contributions and elevating measurement
sensitivity \citep{Hallaji2017,Rostom2022}. 
Beyond these operational
benefits, postselection engenders profound foundational phenomena, notably, the emergence of closed timelike curves within certain quantum
circuit architectures \citep{Lloyd2011,Shepelin2021}. Recent theoretical
and experimental investigations have illuminated the capacity of such
non-classical causal structures to confer metrological advantages that surpass
classical bounds \citep{ArvidssonShukur2023}, opening new avenues
for exploiting temporal quantum correlations in parameter estimation.

Nevertheless, the intrinsic probabilistic nature of postselection and the associated  back-action on the  the meter state introduce
challenges in efficiently compressing and transferring information
from the quantum system to the meter \citep{ArvidssonShukur2020,Yang2024,Yang2024a,Zhong2024,Salvati2024,Zhu2025,PhysRevResearch.6.043084}.
Addressing this challenge requires identifying the optimal operational
parameters that maximize the retention of quantum Fisher information
(QFI) per trial, a central figure of merit quantifying achievable precision
\citep{ArvidssonShukur2020}. 

In this work, we unveil a fundamentally geometric solution to this
problem by demonstrating that the operation on the noncyclic Pancharatnam
phase uniquely optimizes the quality of quantum compression channels. We concentrate on the Pancharatnam phase that emerges naturally from the coherent system-meter interaction, where the parameter-dependent coupling imprints a structural phase relationship on the composite evolution.

The Pancharatnam phase, originally introduced in the context of geometric
phases for polarized light~\citep{Pancharatnam1956Generalized} and
later extended to general quantum evolutions \citep{Berry1987}, prescribes
a phase relation that captures the intrinsic geometry of quantum state
transformations. Its robustness and operational significance have
been verified theoretically and experimentally using interferometric
and polarimetric methods \citep{Larsson2003,Loredo2009,Yakovleva2019,Rostom2023}.
For both pure and mixed states, the Pancharatnam phase manifests as a measurable shift in the interference pattern \citep{Berry1987,Sjoeqvist2000}. Our primary focus is on the Pancharatnam phase as a relative phase difference that can be experimentally accessed and quantified.
Although the Pancharatnam phase has been extensively studied, its role
in optimizing quantum compression channels remains unexplored.

In the following section, we derive the QFI associated with the system\textendash meter
interaction parameter as an explicit function of the channel operator.
We then specify the channel and meter operators, expressing the QFI
accordingly. Building on this foundation, we derive   the Pancharatnam phase that characterizes the quantum
channel and establishes the parallel transport condition necessary
to control the unwanted parallel evolution of the meter state. In the Discussion section, the optimality of the Pancharatnam phase is examined as a natural and physically meaningful parameterization for quantum compression channels. This analysis includes a treatment of general qudit-meter states and provides a clear comparison between scenarios where the Pancharatnam phase is incorporated and those where it is not, emphasizing its crucial role in enhancing compression performance and metrological information encoding.
\vspace{-3mm}
\section{Preliminaries}
\subsection{System\textendash Meter Coupling and Channel Operator }

Consider a quantum system described by the Hilbert space $\mathcal{H}_{s}=\textrm{span}\left\{ |\mathcal{S}_{1}\rangle,|\mathcal{S}_{2}\rangle,\ldots,|\mathcal{S}_{i}\rangle,\ldots\right\} ,$
initially prepared in state $|\mathcal{S}_{i}\rangle$. Suppose
an ancillary system\textemdash the meter\textemdash initially in the
state $|\mathcal{M}_{i}\rangle$. Assuming the system and meter are
initially unentangled, their joint initial state is described by the
tensor product
\begin{equation}
|\Upsilon \rangle=|\mathcal{S}_{i}\rangle\otimes|\mathcal{M}_{i}\rangle.
\end{equation}
Suppose an evolution operator $\hat{\mathds{J}}{\scriptstyle (\lambda,\Theta)}$
acting on the combined system-meter Hilbert space, where $\Theta$
denotes an experimentally tunable postselection parameter controlling
the quantum system, and $\lambda$ is the coupling strength between the system and the meter. The joint state is given by
\begin{equation}
\left|\Upsilon{\scriptstyle (\lambda,\Theta)}\right\rangle =\hat{\mathds{J}}{\scriptstyle (\lambda,\Theta)}|\Upsilon \rangle=\hat{\mathds{J}}{\scriptstyle (\lambda,\Theta)}|\mathcal{S}_{i}\rangle\otimes|\mathcal{M}_{i}\rangle.
\end{equation}
By postselecting the system in the final state $|\mathcal{S}_{f}\rangle$,
the output conditional state becomes
\begin{align}
\left|\Psi{\scriptstyle (\lambda,\Theta)}\right\rangle & \propto\bigl(\left|\!\right.\mathcal{S}_{f}\!\left\rangle \right\langle \!\mathcal{S}_{f}\!\left|\right.\otimes\hat{\mathbb{I}}\bigr)\left|\Upsilon{\scriptstyle (\lambda,\Theta)}\right\rangle\nonumber
\\
& \propto\bigl(\left|\!\right.\mathcal{S}_{f}\!\left\rangle \right\langle \!\mathcal{S}_{f}\!\left|\right.\otimes\hat{\mathbb{I}}\bigr)\hat{\mathds{J}}{\scriptstyle (\lambda,\Theta)}\,\,|\mathcal{S}_{i}\rangle\otimes|\mathcal{M}_{i}\rangle\nonumber \\
 & =|\mathcal{S}_{f}\rangle\otimes\hat{\mathds{K}}{\scriptstyle (\lambda,\Theta)}|\mathcal{M}_{i}\rangle,
\end{align}
where $\hat{\mathds{K}}{\scriptstyle (\lambda,\Theta)}=\left\langle \!\right.\mathcal{S}_{f}\left.\!\right|\hat{\mathds{J}}{\scriptstyle (\lambda,\Theta)}|\mathcal{S}_{i}\rangle$
is the channel operator, representing an effective transformation
acting solely on the meter state. Under the action of $\hat{\mathds{K}}{\scriptstyle (\lambda,\Theta)}$,
the output normalized state 
\begin{equation}
\left|\Psi{\scriptstyle (\lambda,\Theta)}\right\rangle =\mathcal{P}^{-1/2}{\scriptstyle (\lambda,\Theta)}|\mathcal{S}_{f}\rangle\otimes\hat{\mathds{K}}{\scriptstyle (\lambda,\Theta)}|\mathcal{M}_{i}\rangle,\label{eq:state}
\end{equation}
exhibits an amplified sensitivity to weak-coupling strengths $\lambda$.
The normalization function $\mathcal{P}{\scriptstyle (\lambda,\Theta)}$
determines the postselection probability
\begin{align}
\mathcal{P}{\scriptstyle (\lambda,\Theta)} =\left\langle \negmedspace\right.\mathcal{M}_{i}\left.\!\right|\hat{\mathds{K}}^{\dagger}{\scriptstyle (\lambda,\Theta)}\hat{\mathds{K}}{\scriptstyle (\lambda,\Theta)}|\mathcal{M}_{i}\rangle=\left\langle \!\right.\hat{\mathds{K}}^{\dagger}{\scriptstyle (\lambda,\Theta)}\hat{\mathds{K}}{\scriptstyle (\lambda,\Theta)}\left.\!\right\rangle. 
\end{align}

\subsection{QFI: Total and Parallel Evolution Characterized by $\hat{\mathds{K}}{\scriptstyle (\lambda,\Theta)}$}

In this subsection, we analyze the total and parallel contributions to
the QFI as expressed through the operator $\hat{\mathds{K}}{\scriptstyle (\lambda,\Theta)}$ 

QFI quantifies the sensitivity of the quantum state $|\Psi{\scriptstyle (\lambda,\Theta)}\rangle$
to changes in the parameter $\lambda$ and sets the theoretical limits
for estimation precision via the quantum Cram\'{e}r-Rao bound \cite{ArvidssonShukur2020}
\begin{equation}
\textrm{\textbf{Var}}(\lambda)\geq\frac{1}{M\cdot\mathcal{I}^{\perp}{\scriptstyle (\lambda,\Theta)}},\label{eq:=000020Cram=0000E9r-Rao=000020bound}
\end{equation}
where M is the number of measurements and $\mathcal{I}^{\perp}{\scriptstyle (\lambda,\Theta)}$
is the observed QFI of a single measurement. QFI is a gauge-invariant
quantity (Appendix \ref{Gauge invariance}), as it depends only on the orthogonal component $|\partial_{\lambda}\Psi{\scriptstyle (\lambda,\Theta)}\rangle^{\perp}$
(Appendix \ref{subsec:Geometrical-description-of}) 
\begin{align}
\mathcal{I}^{\perp}{\scriptstyle (\lambda,\Theta)}= & 4\bigr\Vert|\partial_{\lambda}\Psi{\scriptstyle (\lambda,\Theta)}\rangle^{\perp}\bigr\Vert^{2}\nonumber \\
= & 4\bigl(\langle\partial_{\lambda}\Psi{\scriptstyle (\lambda,\Theta)}|\partial_{\lambda}\Psi{\scriptstyle (\lambda,\Theta)}\rangle-|\langle\Psi{\scriptstyle (\lambda,\Theta)}|\partial_{\lambda}\Psi{\scriptstyle (\lambda,\Theta)}\rangle|^{2}\bigr)\nonumber\\
= & 4\bigl(\bigr\Vert|\partial_{\lambda}\Psi{\scriptstyle (\lambda,\Theta)}\rangle\bigr\Vert^{2}-\bigr\Vert|\partial_{\lambda}\Psi{\scriptstyle (\lambda,\Theta)}\rangle^{\parallel}\bigr\Vert^{2}\bigr)\label{eq:standard=00003D000020qfi}
\end{align}
The term $\bigr\Vert|\partial_{\lambda}\Psi{\scriptstyle (\lambda,\Theta)}\rangle\bigr\Vert^{2}$
quantifies   the total rate of change of the quantum state $|\Psi{\scriptstyle (\lambda,\Theta)}\rangle$
in Hilbert space with respect to the parameter $\lambda$. The second
term corresponds to the parallel component $\bigr\Vert|\partial_{\lambda}\Psi{\scriptstyle (\lambda,\Theta)}\rangle^{\parallel}\bigr\Vert^{2}$\textemdash a
gauge-dependent contribution and thus does not contribute to physically
accessible information. 

By substituting the postselected state $\left|\Psi{\scriptstyle (\lambda,\Theta)}\right\rangle $,
the QFI under postselection becomes (Appendix \ref{sec:Derivation-of-the})
\begin{align}
\mathcal{I}^{\perp}{\scriptstyle (\lambda,\Theta)} & =\,\,4\mathcal{P}^{-2}{\scriptstyle (\lambda,\Theta)}\bigl[\mathcal{Q}^{T}{\scriptstyle (\lambda,\Theta)}\mathcal{P}_{\lambda}{\scriptstyle (\lambda,\Theta)}-\mathcal{Q}^{\parallel}{\scriptstyle (\lambda,\Theta)}\bigr]\nonumber \\
 & =\mathcal{I}^{T}{\scriptstyle (\lambda,\Theta)}-\mathcal{I}^{\parallel}{\scriptstyle (\lambda,\Theta)}\label{postselected QFI},
\end{align}
where $\mathcal{Q}^{T}{\scriptstyle (\lambda,\Theta)}=\left\langle \!\right.\partial_{\lambda}\hat{\mathds{K}}^{\dagger}{\scriptstyle (\lambda,\Theta)}\partial_{\lambda}\hat{\mathds{K}}{\scriptstyle (\lambda,\Theta)}\left.\!\right\rangle $
is the average total change in the channel operator with respect to
the parameter $\lambda$, and $\mathcal{Q}^{\parallel}{\scriptstyle (\lambda,\Theta)}=\left|\negthickspace\right.\left\langle \!\right.\hat{\mathds{K}}^{\dagger}{\scriptstyle (\lambda,\Theta)}\,\partial_{\lambda}\hat{\mathds{K}}{\scriptstyle (\lambda,\Theta)}\left.\!\right\rangle \left.\negthickspace\right|^{2}$
is its parallel evolution. The parallel component here is not simply
subtracted as a gauge artifact but interacts nontrivially with $\mathcal{P}{\scriptstyle (\lambda,\Theta)}$. 

The recent theory of compression quantum channels \citep{Yang2024} establishes necessary
and sufficient conditions for a postselection channel to be lossless,
meaning the average QFI of the retained outcomes equals that of the
standard QFI (i.e., without postselection), thereby preserving metrological sensitivity despite
compression. This condition, detailed in Table~I, Theorem~1, and
supplementary materials of the ref. \citep{Yang2024}, is mathematically equivalent
to the condition $\mathcal{Q}^{\parallel}{\scriptstyle (\lambda,\Theta)}=0$
in this work.

We demonstrate in Sections \ref{Controlling Parallel Evolution} and \ref{Complete Nullification} that when the interaction induces Pancharatnam
phase shifts in the post-selected system, the parallel component $\mathcal{Q}^{\parallel}{\scriptstyle (\lambda,\Theta)}$
exhibits phase-dependent interference controlled by the parameter
$\Theta$. This interference allows precise tuning and  complete suppression
of $\mathcal{Q}^{\parallel}{\scriptstyle (\lambda,\Theta)}$. Conversely, in the absence of Pancharatnam phase shifts, $\mathcal{Q}^{\parallel}{\scriptstyle (\lambda,\Theta)}$
reduces to a nonvanishing, $\Theta$-independent squared magnitude
of the expectation derivative. This critical difference in $\Theta$-sensitive
coherence effects in the former case are essential for enhanced quantum
parameter estimation control.

\subsection{QFI Yield per Trial}

In postselected quantum metrology, the number of successful postselections
is given by $M=\mathcal{N}\mathcal{P}{\scriptstyle (\lambda,\Theta)}$,
where $\mathcal{N}$ denotes the total number of experimental trials
and accounts for both successful and unsuccessful postselection attempts. The achievable precision reads
\begin{align}
\textrm{\textrm{\textbf{Var}}(\ensuremath{\lambda})} & \geq\frac{1}{M\mathcal{I}^{\perp}{\scriptstyle (\lambda,\Theta)}}=\frac{1}{\mathcal{N}\mathcal{P}{\scriptstyle (\lambda,\Theta)}\mathcal{I}^{\perp}{\scriptstyle (\lambda,\Theta)}} =\frac{1}{\mathcal{N}\mathcal{T}{\scriptstyle (\lambda,\Theta)}},
\label{var with T}
\end{align}
and the quantum advantage gained via postselection is quantified by the
total QFI per trial \citep{Giovannetti2011Advances,ArvidssonShukur2020}
\begin{equation}
\mathcal{T}{\scriptstyle (\lambda,\Theta)}=\mathcal{P}{\scriptstyle (\lambda,\Theta)}\mathcal{I}^{\perp}{\scriptstyle (\lambda,\Theta)},
\end{equation}
which serves as the central metric for evaluating the efficiency of
the compression channel. This metric balances high precision\textemdash potentially
large $\mathcal{I}^{\perp}{\scriptstyle (\lambda,\Theta)}$\textemdash against
the typically reduced success probability $\mathcal{P}{\scriptstyle (\lambda,\Theta)}$
arising from postselection on rare events \footnote{Under weak interaction, the system and meter become non-maximally
entangled, a state expected to suppress the quantum advantage typically
offered by maximally entangled states \citep{Wu2025}.}. Specifically, when $\mathcal{T}{\scriptstyle (\lambda,\Theta)}$
scales proportionally with the   square of the total number of  probes,
the metrological protocol attains the Heisenberg limit for precision \citep{Giovannetti2005}.

Our main objective is  to identify the optimal postselection parameter
$\Theta_{\parallel}$ that satisfies $\mathcal{Q}^{\parallel}{\scriptstyle }=0$,
where all parameter-induced changes are encoded fully in physically
distinguishable directions in Hilbert space. This is a necessary condition
for maximizing $\mathcal{T}{\scriptstyle (\lambda,\Theta)}$ as we will demonstrate below.

\section{Determination of the channel operator $\hat{\mathds{K}}{\scriptstyle (\lambda,\Theta)}$}

Consider a  quantum system described by a two-mode
Hilbert space, whose structure can be elegantly and compactly captured
through the Jordan\textendash Schwinger map \citep{Kok2010}
\begin{equation}
\begin{gathered}\hat{J}_{x}=\frac{1}{2}(\hat{a}_{1}^{\dagger}\hat{a}_{2}+\hat{a}_{2}^{\dagger}\hat{a}_{1}),~~\hat{J}_{y}=\frac{i}{2}(\hat{a}_{2}^{\dagger}\hat{a}_{1}-\hat{a}_{1}^{\dagger}\hat{a}_{2}),\\
\hat{J}_{z}=\frac{1}{2}(\hat{a}_{1}^{\dagger}\hat{a}_{1}-\hat{a}_{2}^{\dagger}\hat{a}_{2}),~~\hat{J}_{t}=\frac{1}{2}(\hat{a}_{1}^{\dagger}\hat{a}_{1}+\hat{a}_{2}^{\dagger}\hat{a}_{2}).
\end{gathered}
\end{equation}
where ($\hat{a}_{1}^{\dagger}$, $\hat{a}_{2}^{\dagger}$) and ($\hat{a}_{1}$,
$\hat{a}_{2}$) are Boson creation and annihilation operators for modes 1 and 2, respectively. These operators
form a closed algebra under the commutator bracket, satisfying $[\hat{J}_{i},\hat{J}_{j}]=\imath\epsilon_{ijk}\hat{J}_{k},$
where $\epsilon_{ijk}$ is the Levi-Civita symbol, and $\hat{J}_{t}$
commutes with all other $\hat{J}_{i}$. 

The system\textendash meter interaction is commonly modeled by a Hamiltonian
that couples the meter to one or more system observables. The total
Hilbert space for the combined system is given by the tensor product
$\mathcal{H}=\mathcal{H}_{\mathcal{S}}\otimes\mathcal{H}_{\mathcal{M}}$,
where $\mathcal{H}_{\mathcal{S}}$ and $\mathcal{H}_{\mathcal{M}}$
denote the Hilbert spaces of the system and the meter, respectively.
The interaction dynamics can be described by the unitary operator
\begin{equation}
\hat{\mathcal{\varLambda}}{\scriptstyle (\lambda)}=e^{ \imath\lambda(\hat{J}_{t}^{(\mathcal{S})}+\hat{J}_{z}^{(\mathcal{S})})\otimes\,\hat{\mathcal{M}}},\label{eq:Interaction=000020unitary=000020operator}
\end{equation}
where $\hat{\mathcal{M}}$ is a Hermitian operator
acting on the meter's state. The operator $\hat{\mathcal{\varLambda}}{\scriptstyle (\lambda)}$
encapsulates the interaction between the system\textquoteright s mode
1 and the meter \footnote{The structure of the interaction operator $\hat{\mathcal{\varLambda}}{\scriptstyle (\lambda)}$ considered in Eq.~\eqref{eq:Interaction=000020unitary=000020operator} is not unique. Alternative forms consistent with the underlying physical principles can be employed without altering the core operational framework. The essential criterion, as we demonstrate,
is to select a meter operator that induces a Pancharatnam phase shift,
manifesting as a relative phase in the interference pattern. The choice
of system operator is guided by experimental feasibility and can be
adapted to diverse platforms, including gravitationally induced entanglement \cite{Nguyen2020}, controlled-Z gates \cite{Kok2007}, cross-Kerr
nonlinearities \cite{Rostom_2020} and one- and two-axis twisting spin-squeezed states \cite{byrnes2021quantum}.
For instance, selecting an interaction operator such as $\hat{\mathcal{\varLambda}}{\scriptstyle (\lambda)}=e^{i\lambda\hat{J}_{z}^{(\mathcal{S})}\otimes\,\hat{\mathcal{M}}},$
yields $\hat{\mathds{J}}{\scriptstyle (\lambda,\Theta)}=e^{i\hat{J}_{y}^{(\mathcal{S})}\otimes\bigl(\Theta\hat{\mathbb{I}}-\lambda\hat{\mathcal{\mathcal{M}}}\bigr)}$,
preserving the Pancharatnam phase shift and the interpretation of
the results. While this choice may alter the behavior of $\mathcal{Q}^{T}{\scriptstyle (\lambda,\Theta)}$
and $\mathcal{Q}^{\parallel}{\scriptstyle (\lambda,\Theta)}$, the
fundamental operational principle remains unchanged. }. Incorporating a controlled phase applied to mode
2, the overall operator governing the system\textquoteright s evolution
is then defined as
\begin{align}
\hat{\mathds{J}}{\scriptstyle (\lambda,\Theta)} &=\bigl [e^{-\imath\frac{\pi}{2}\hat{J}_{x}^{(\mathcal{S})}}\otimes\hat{\mathbb{I}}\bigr]\,\bigl[e^{\imath\Theta(\hat{J}_{t}^{(\mathcal{S})}-\hat{J}_{z}^{(\mathcal{S})})}\otimes\hat{\mathbb{I}}\bigr]\nonumber \\
 & \hspace{0.4cm}\times\hat{\mathcal{\varLambda}}{\scriptstyle (\lambda)}\,\bigl[e^{+\imath\frac{\pi}{2}\hat{J}_{x}^{(\mathcal{S})}}\otimes\hat{\mathbb{I}}\bigr].
\end{align}
The operator $e^{\pm \imath\frac{\pi}{2}\hat{J}_{x}^{(\mathcal{S})}}$ generates a
rotation about the x-axis, effectively implementing the 50:50 splitting
and recombination of the two quantum modes. Using the identity $\hat{U}e^{\hat{A}}\hat{U}^{\dagger}=e^{\hat{U}\hat{A}\hat{U}^{\dagger}}$,
and Baker-Campbell-Haussdorff relations $\ensuremath{e^{-\imath\frac{\pi}{2}\hat{J}_{x}^{(\mathcal{S})}}\hat{J}_{z}^{(\mathcal{S})}e^{+\imath\frac{\pi}{2}\hat{J}_{x}^{(\mathcal{S})}}=-\hat{J}_{y}^{(\mathcal{S})},\,\,e^{-\imath\frac{\pi}{2}\hat{J}_{x}^{(\mathcal{S})}}\hat{J}_{t}^{(\mathcal{S})}e^{+\imath\frac{\pi}{2}\hat{J}_{x}^{(\mathcal{S})}}=\hat{J}_{t}^{(\mathcal{S})}}$,
the operator $\hat{\mathds{J}}{\scriptstyle (\lambda,\Theta)}$
can be expressed as 
\begin{equation}
\hat{\mathds{J}}{\scriptstyle (\lambda,\Theta)} =\bigl[e^{\imath\Theta(\hat{J}_{t}^{(\mathcal{S})}+\hat{J}_{y}^{(\mathcal{S})})}\otimes\hat{\mathbb{I}}\bigr]e^{\imath\lambda(\hat{J}_{t}^{(\mathcal{S})}-\hat{J}_{y}^{(\mathcal{S})})\otimes\,\hat{\mathcal{M}}}.
\label{EQ14}
\end{equation}

 The pre- and postselected states can be chosen as ($|\mathcal{S}_{i}\rangle=|j,m_{i}\rangle$
and $|\mathcal{S}_{f}\rangle=|j,m_{f}\rangle$), where \citep{Pezze2018}
\[
\ensuremath{|j,m\rangle\equiv|n_{1},n_{2}\rangle=\frac{(a_{1}^{\dagger})^{j+m}(a_{2}^{\dagger})^{j-m}}{\sqrt{(j+m)!(j-m)!}}|0,0\rangle,}
\]
where $j=\frac{n_{1}+n_{2}}{2}$ is the total angular momentum quantum number,
$m=\frac{n_{1}-n_{2}}{2}$ is the eigenvalue of $\hat{J}_{z}$ (the magnetic
quantum number), and $|0,0\rangle$ is the vacuum state. Consequently, 
the operator $\hat{\mathds{K}}{\scriptstyle (\lambda,\Theta)}$,  can be obtained by projecting $\hat{\mathds{J}}{\scriptstyle (\lambda,\Theta)}$
onto the pre- and postselected states, 
yielding 
\begin{align}
\hat{\mathds{K}}_{m_{f},m_{i}}^{(j)}{\scriptstyle (\lambda,\Theta)}= & \left\langle \!\right.\mathcal{S}_{f}\left.\!\right|\hat{\mathds{J}}{\scriptstyle (\lambda,\Theta)}|\mathcal{S}_{i}\rangle\nonumber \\
= & \left\langle \!\right.j,m_{f}\left.\!\right|\hat{\mathds{J}}{\scriptstyle (\lambda,\Theta)}|j,m_{i}\rangle\nonumber \\
= & e^{\imath j\Theta}e^{\imath j\lambda\hat{\mathcal{M}}}\mathcal{\mathcal{\hat{F}}}_{m_{f},m_{i}}^{(j)}{\scriptstyle (\lambda,\Theta)}.\label{eq:compression=000020channel=000020operator}
\end{align}
where  $\hat{J}_{t}^{(\mathcal{S})}|j,m_{i}\rangle=j|j,m_{i}\rangle$,   $\hat{J}_{z}^{(\mathcal{S})}|j,m\rangle=m|j,m\rangle$ and
\begin{equation}
\mathcal{\hat{F}}_{m_{f},m_{i}}^{(j)}{\scriptstyle (\lambda,\Theta)}=\langle j,m_{f}|e^{\imath\hat{J}_{y}^{(\mathcal{S})}\otimes(\Theta\hat{\mathbb{I}}-\lambda\hat{\mathcal{M}})}|j,m_{i}\rangle.
\label{EQ16}
\end{equation}
 Applying $\mathcal{\hat{F}}_{m_{f},m_{i}}^{(j)}{\scriptstyle (\lambda,\Theta)}$ on the meter  eigenstate $|b_{k}\rangle$ of the operator $\hat{\mathcal{M}}$
gives 
\[
\mathcal{\hat{F}}_{m_{f},m_{i}}^{(j)}{\scriptstyle (\lambda,\Theta)}|b_{k}\rangle=d_{m_{f},m_{i}}^{(j)}{({ \Theta-b_{k}\lambda})}|b_{k}\rangle,
\]
where $d_{m_{f},m_{i}}^{(j)}{ { ({ \Theta-b_{k}\lambda})}}$
is the Wigner d-matrix element \cite{varshalovich1988} that corresponds to the eigenvalues $b_{k}$
of the meter operator $\hat{\mathcal{M}}$.

Note that the  operator in Eq. \eqref{EQ14} can be equivalently expressed as
\begin{equation}
\hat{\mathds{J}}{\scriptstyle (\lambda,\Theta)}=e^{\imath\hat{J}_{t}^{(\mathcal{S})}\otimes(\Theta\hat{\mathbb{I}}+\lambda\hat{\mathcal{\mathcal{M}}})}e^{\imath\hat{J}_{y}^{(\mathcal{S})}\otimes(\Theta\hat{\mathbb{I}}-\lambda\hat{\mathcal{\mathcal{M}}})}.
\label{EQ17}
\end{equation}
Thus $\hat{\mathcal{F}}_{m_{f},m_{i}}^{(j)}{\scriptstyle (\lambda,\Theta)}$ (Eqs. \eqref{eq:compression=000020channel=000020operator} and \eqref{EQ16})
reveals that the post-selection probability is primarily governed
by the second term in Eq. \eqref{EQ17}, which incorporates contributions
from both the post-selection parameter $\Theta$ and interaction strength
$\lambda$. The first term in Eq. \eqref{eq:compression=000020channel=000020operator} shows that $e^{\imath\Theta\hat{J}_{t}^{(\mathcal{S})}}$
from \eqref{EQ17} induces only a trivial global phase. Conversely, the
second term in Eq. \eqref{eq:compression=000020channel=000020operator} demonstrates that $e^{\imath\lambda\hat{J}_{t}^{(\mathcal{S})}\otimes\hat{\mathcal{M}}}$ from \eqref{EQ17} 
imparts a non-trivial rotation to the post-selected state, which can
impact the QFI in general scenarios.

From now on and for simplicity, the explicit dependence on the parameters
$\lambda$ and $\Theta$ will be omitted in the notation of operators
and functions whenever no ambiguity arises.

\subsection{General Meter Operators}

Here, we present a mathematical formula that facilitates
the determination of the channel operator $\hat{\mathds{K}}_{m_{f},m_{i}}^{(j)}$
for an arbitrary meter state.

The states $|j,m_{i}\rangle$ and $|j,m_{f}\rangle$ considered above are eigenstates of
$\hat{J}_{z}^{(\mathcal{S})}$, and the operator   $\hat{\mathcal{M}}$ acts solely on the meter state. To evaluate the matrix element $\mathcal{\mathcal{\hat{F}}}_{m_{f},m_{i}}^{(j)}$, 
we first express the system states in the eigenbasis of $\hat{J}_{y}^{(\mathcal{S})}$
using the Wigner d-matrix
\begin{equation}
|j,m\rangle=\sum_{m_{y}=-j}^{j}d_{m_{y},m}^{(j)}(\pi/2)|j,m_{y}\rangle,
\end{equation}
where $d_{m_{y},m}^{(j)}(\pi/2)$ is the Wigner d-matrix element
for a rotation by angle $\pi/2$ about the $y$-axis. The operator
\begin{equation}
e^{\imath \hat{J}_{y}^{(\mathcal{S})}\otimes(\Theta\,\hat{\mathbb{I}}-\lambda\,\hat{\mathcal{M}})}
\end{equation}
acts diagonally in this basis, yielding eigenvalues $m_{y}$ for $\hat{J}_{y}^{(\mathcal{S})}$,
so that its action on $|j,m_{y}\rangle$ is simply a multiplication
by $\exp[\imath\,m_{y}(\Theta\,\hat{\mathbb{I}}-\lambda\,\hat{\mathcal{M}})]$.
Substituting the expansions for both $|j,m_{i}\rangle$ and $|j,m_{f}\rangle$,
and using the orthogonality of the $\hat{J}_{y}$ eigenstates, the
matrix element reduces to a sum over products of Wigner d-matrix
elements and the exponential operator
\begin{align}
\mathcal{\mathcal{\hat{F}}}_{m_{f},m_{i}}^{(j)}=  \sum_{m_{y}=-j}^{j}d_{m_{y},m_{f}}^{(j)}(\frac{\pi}{2})d_{m_{y},m_{i}}^{(j)}(\frac{\pi}{2})  e^{\imath m_{y}(\Theta\,\hat{\mathbb{I}}-\lambda\,\hat{\mathcal{M}})},\label{eq:matrix=000020element=000020F}
\end{align}
where the sum runs over all eigenvalues $m_{y}$ of $\hat{J}_{y}^{(\mathcal{S})}$.
This construction relates the matrix element $\mathcal{\mathcal{\hat{F}}}_{m_{f},m_{i}}^{(j)}$
to a weighted sum over all possible projections onto the $\hat{J}_{y}$
eigenbasis, with each term governed by the system's rotation properties
and the meter operator.

\subsection{The Pancharatnam
Phase} \label{The Pancharatnam
Phase}

In this section, we demonstrate that, for a general meter state, the
postselected system operates as an interferometric setup exhibiting
characteristic Pancharatnam phase effects.

From Eq. (\ref{eq:matrix=000020element=000020F}),
the matrix element $\mathcal{F}_{m_{k},m_{i}}^{(j)}$
can be written as a finite sum of exponentials in $\Theta$ and $\hat{\mathcal{M}}$ for
any $j$. By pre- and postselecting the highest and
lowest weight states  denoted by $|j,m_{i}\rangle=|j,j\rangle$,
and $|j,m_{f}\rangle=|j,-j\rangle$, respectively, $\Delta m=m_{i}-m_{f}=2j$  and the channel operator (\ref{eq:compression=000020channel=000020operator})
yields
\begin{align}
\hat{\mathds{K}}_{-j,j}^{(j)}&=  e^{\imath j\Theta}e^{\imath j\lambda\hat{\mathcal{M}}}\mathcal{\mathcal{\hat{F}}}_{-j,j}^{(j)}\nonumber \\
&=  (-1)^{j}e^{\imath j\Theta}e^{\imath j\lambda\hat{\mathcal{M}}}\sin^{2j}(\frac{\Theta\hat{\mathbb{I}}-\lambda\hat{\mathcal{M}}}{2})\nonumber \\
&=  \frac{(-1)^{j}}{(2i)^{2j}}(e^{\imath\Theta}\hat{\mathbb{I}}-\hat{\mathcal{O}}_{\lambda})^{2j},
\end{align}
where $\hat{\mathcal{O}}_{\lambda}=e^{\imath\lambda\hat{\mathcal{M}}}$.

Since $\hat{\mathcal{O}}_{\lambda}\text{ is unitary, then }\hat{\mathcal{O}}_{\lambda}^{\dagger}\hat{\mathcal{O}}_{\lambda}=\hat{\mathbb{I}},\mathrm{~}\langle\hat{\mathcal{O}}_{\lambda}^{\dagger}\hat{\mathcal{O}}_{\lambda}\rangle=1,$
and its expectation value satisfy: $\left\langle \!\right.\hat{\mathcal{O}}_{\lambda}^{\dagger}\left.\!\right\rangle =\left\langle \!\right.\hat{\mathcal{O}}_{\lambda}\left.\!\right\rangle ^{*}.$
Define $\left\langle \!\right.\hat{\mathcal{O}}_{\lambda}\left.\!\right\rangle =\left|\negthickspace\right.\left\langle \!\right.\hat{\mathcal{O}}_{\lambda}\left.\!\right\rangle \left.\negthickspace\right|e^{\imath\textrm{Im\,ln}\left\langle \!\right.\hat{\mathcal{O}}_{\lambda}\left.\!\right\rangle }.$
The postselection probability $\mathcal{P}_{m_{f},m_{i}}^{(j)}=\mathcal{P}_{-j,j}^{(j)}$ for $j=1/2$ can be expressed as
\begin{align}
\mathcal{P}_{-\frac{1}{2},\frac{1}{2}}^{(\frac{1}{2})}= & \tfrac{1}{4}(2-e^{\imath\Theta}\left\langle \!\right.\hat{\mathcal{O}}^{\dagger}_{\lambda}\left.\!\right\rangle -e^{-\imath\Theta}\left\langle \!\right.\hat{\mathcal{O}}_{\lambda}\left.\!\right\rangle )\nonumber \\
= & \tfrac{1}{2}\bigl[1-\bigl|\negthickspace\bigr.\left\langle \!\right.\hat{\mathcal{O}}_{\lambda}\left.\!\right\rangle \bigl.\negthickspace\bigr|\cos(\Theta-\textrm{Im\,ln}\left\langle \!\right.\hat{\mathcal{O}}_{\lambda}\left.\!\right\rangle )\bigr],\label{eq:pancharatnam=000020shift}
\end{align}
where $\left|\negthickspace\right.\left\langle \!\right.\hat{\mathcal{O}}_{\lambda}\left.\!\right\rangle \left.\negthickspace\right|$
is the visibility, and $\textrm{Im\,ln}\left\langle \!\right.\hat{\mathcal{O}}_{\lambda}\left.\!\right\rangle $
is Pancharatnam phase difference between the two system's modes \footnote{Here, the meter can be interpreted as an environmental degree of freedom
interacting with the system \citep{Sjoeqvist2000,Yakovleva2019a}.}, see Appendix \ref{sec:Pancharatnam-phase} for more details on the
interpretation of this phase.  It can be defined
to be the phase by which one mode must be retarded or advanced to
make the postselection probability in the system reaches its minimum \citep{Pancharatnam1956Generalized}.

For higher values of $j$, the postselection probability is given by 
\begin{align}
\mathcal{P}_{-j,j}^{(j)}= & \frac{1}{16^{j}}\left\langle \!\right.(2\hat{\mathbb{I}}-e^{\imath\Theta}\hat{\mathcal{O}}_{\lambda}^{\dagger}-e^{-\imath\Theta}\hat{\mathcal{O}}_{\lambda})^{2j}\left.\!\right\rangle ,
\end{align}
which represents a $2j$-th order interference kernel that generates
quantum fringes with phase sensitivity controlled by $\Theta$, phase
shift inherited from the eigenvalues of $\hat{\mathcal{O}}_{\lambda}$,
and contrast scaling determined by $j$, see Fig. \ref{fig:0}(a).
It is crucial to emphasize that the Pancharatnam phase $\textrm{Im\,ln}\left\langle \!\right.\hat{\mathcal{O}}_{\lambda}\left.\!\right\rangle $ remains invariant with respect to $j$, \(m_i\) and \(m_f\), see Fig.~\ref{fig:0}.

However, due to the exponential dependence of the postselection probability
on $\Delta m=2j$, the optimal postselection---which can be viewed
as a lossless compression channel---for higher values
of $j$ $(j>1/2)$ can be realized when the transition between pre-
and postselected states corresponds exactly to a single quantum $\Delta m=1$.
A natural and effective choice fulfilling this condition is $m_{i}=j$
and $m_{f}=j-1$, which corresponds to
\begin{equation}
\mathcal{\mathcal{\hat{F}}}_{j-1,j}^{(j)}=-\sqrt{2j}(\cos\frac{\Theta\hat{\mathbb{I}}-\lambda\hat{\mathcal{M}}}{2})^{2j-1}\sin\frac{\Theta\hat{\mathbb{I}}-\lambda\hat{\mathcal{M}}}{2}.\label{eq:F=000020for=000020j,=000020and=000020j-1}
\end{equation}
For $j=1$, the channel operator takes the form
\begin{align}
\hat{\mathds{K}}_{0,1}^{(1)}= & \frac{-1}{\sqrt{2}}\bigl(e^{\imath 2\Theta}\hat{\mathbb{I}}-e^{\imath 2\lambda\hat{\mathcal{M}}}\bigr),
\end{align}
and the corresponding postselection probability is given by
\begin{align}
\mathcal{P}_{0,1}^{(1)}= & \tfrac{1}{4}\bigl[1-\left|\negthickspace\right.\left\langle \!\right.\hat{\mathcal{O}}_{\lambda}^{2}\left.\!\right\rangle \left.\negthickspace\right|\cos2\bigl(\Theta-\textrm{Im\,ln}\left\langle \!\right.\hat{\mathcal{O}}_{\lambda}\left.\!\right\rangle \bigr)\bigr]
\end{align}
where $\hat{\mathcal{O}}_{\lambda}$ is diagonal in the meter basis
and hence $\textrm{Im\,ln}\left\langle \!\right.\hat{\mathcal{O}}_{\lambda}^{2}\left.\!\right\rangle =2\textrm{Im\,ln}\left\langle \!\right.\hat{\mathcal{O}}_{\lambda}\left.\!\right\rangle $.
The oscillatory behavior of $\mathcal{P}_{0,1}^{(1)}$
manifests at twice the frequency of $\mathcal{P}_{-\frac{1}{2},\frac{1}{2}}^{(\frac{1}{2})}$,
reflecting a harmonic doubling in the interference pattern, while
both retain identical Pancharatnam phase shifts. 

For higher values of
$j$, analytical treatment becomes increasingly intricate due to the
complexities involved in handling powers of operator-valued functions.
Nevertheless, numerical simulations, such as those illustrated in
Fig. \ref{fig:0},  confirm the invariance of the Pancharatnam phase
with respect to variations in $j,m_{i}$ and $m_{f}$.

It is important to note here that alternative pre- and post-selected
states $m_{i}$ and $m_{f}$ can be  explored. For instance,
with $j=2$, choosing $m_{i}=1$ and $m_{f}=0$, entails a single-quantum
transition governed by the channel operator
\begin{align}
\hat{\mathds{K}}_{0,1}^{(2)}=\sqrt{\frac{3}{8}} & \bigl(e^{\imath 4\Theta}\hat{\mathbb{I}}-\hat{\mathcal{O}}_{\lambda}^{4}\bigr),
\end{align}
which likewise induces the identical Pancharatnam phase shift as in Eq. \eqref{eq:pancharatnam=000020shift}, while
manifesting more rapid oscillatory behavior.

\subsection{Controlling Parallel Evolution $\mathcal{Q}^{\parallel}$ }\label{Controlling Parallel Evolution} 

If the postselected quantum system acquires a noncyclic Pancharatnam
phase characterized by $\textrm{Im}\ln\langle\hat{\mathcal{O}}_{\lambda}\rangle\neq0$,
the geometric connection   $\langle\hat{\mathcal{O}}_{\lambda}^{\dagger}\partial_{\lambda}\hat{\mathcal{O}}_{\lambda}\rangle$
must remain nonzero (see Appendices \ref{sec:Quantum-parallel-transport} and \ref{sec:Pancharatnam-phase}). This nonvanishing connection enables precise control over the parallel term
contribution to the QFI via the parameter $\Theta$, thereby enabling
the realization of a lossless quantum compression channel.

To illustrate this optimization mechanism, the QFI can be conveniently
reformulated in terms of the meter\textquoteright s interaction operators
$\hat{\mathcal{O}}_{\lambda}$. 
For instance, when $j=1/2$, the first
term, which corresponds to the total rate of change, can be expressed
as
\begin{equation}
\mathcal{Q}^{T}=\left\langle \!\right.\partial_{\lambda}(\hat{\mathds{K}}_{-\frac{1}{2},\frac{1}{2}}^{(\frac{1}{2})})^{\dagger}\partial_{\lambda}\hat{\mathds{K}}_{-\frac{1}{2},\frac{1}{2}}^{(\frac{1}{2})}\left.\!\right\rangle=\tfrac{1}{4}\left\langle \!\right.\partial_{\lambda}\hat{\mathcal{O}}_{\lambda}^{\dagger}\partial_{\lambda}\hat{\mathcal{O}}_{\lambda}\left.\!\right\rangle .\label{eq:total2}
\end{equation}
At the same time, the parallel term   becomes
\begin{align}
\mathcal{Q}^{\parallel} & =\left|\negthickspace\right.\left\langle \!\right.(\hat{\mathds{K}}_{-\frac{1}{2},\frac{1}{2}}^{(\frac{1}{2})})^{\dagger}\partial_{\lambda}\hat{\mathds{K}}_{-\frac{1}{2},\frac{1}{2}}^{(\frac{1}{2})}\left.\negthickspace\right|^{2}\nonumber\\
 & =\tfrac{1}{16}\bigl|\left\langle \!\right.\partial_{\lambda}\hat{\mathcal{O}}_{\lambda}\left.\!\right\rangle -e^{\imath\Theta}\left\langle \!\right.\hat{\mathcal{O}}_{\lambda}^{\dagger}\partial_{\lambda}\hat{\mathcal{O}}_{\lambda}\left.\!\right\rangle \bigr|^{2}.\label{eq:q2=000020parallel}
\end{align}
When the connection  $\langle\hat{\mathcal{O}}_{\lambda}^{\dagger}\partial_{\lambda}\hat{\mathcal{O}}_{\lambda}\rangle\neq0$, active suppression
of the parallel term $\mathcal{Q}^{\parallel}{\scriptstyle (\Theta)}$ becomes
possible: by tuning $\Theta$ to align with the operator's intrinsic
phase, $\mathcal{Q}^{\parallel}{\scriptstyle (\Theta)}$
can be driven to zero, thus surpassing the precision achievable under
parallel transport $\langle\hat{\mathcal{O}}_{\lambda}^{\dagger}\partial_{\lambda}\hat{\mathcal{O}}_{\lambda}\rangle=0$.

Conversely, when the connection vanishes 
$\langle\hat{\mathcal{O}}_{\lambda}^{\dagger}\partial_{\lambda}\hat{\mathcal{O}}_{\lambda}\rangle=0,$
$\mathcal{Q}^{\parallel}{\scriptstyle (\Theta)}$ simplifies
to $|\langle\partial_{\lambda}\hat{\mathcal{O}}_{\lambda}\rangle|^{2}$,
reflecting orthogonal fluctuations determined by the magnitude of
the derivative and unaffected by $\Theta$.

The extension to higher values of $j$ (particularly, $\hat{\mathds{K}}_{0,1}^{(1)}$ and $\hat{\mathds{K}}_{0,1}^{(2)}$) proceeds analogously to the framework established in the preceding section. In Section \ref{Complete Nullification}, we demonstrate that for any value of $j$, the parallel
component $\mathcal{Q}^{\parallel}$ can be fully controlled and nullified
through the same value of the parameter $\Theta$, especially when the pre- and postselection
processes induce transitions involving a single quantum change (lossless
quantum channel).

\subsection{Qudit Meter State\label{Qudit Meter State}}

Additionally, we consider a specific yet sufficiently general qudit
meter state $\ensuremath{|\mathcal{M}_{i}\rangle=|\mathrm{\mathcal{M}}_{d}^{(n)}\rangle=\frac{1}{\sqrt{d}}\sum_{k=0}^{d-1}|b_{k}\rangle^{\otimes n}}$,
for $n$ subsystems and $d$-dimensional Hilbert space spanned by the symmetric subspace $\mathcal{B}=\{|b_{0}\rangle^{\otimes n}$,
$|b_{1}\rangle^{\otimes n}$, ..., $|b_{d-1}\rangle^{\otimes n}\}$.

The meter operator can be defined as
\begin{equation}
e^{\imath n\lambda\hat{\mathcal{M}}}=\sum_{k=0}^{d-1}e^{\imath n b_{k}\lambda}(|b_{k}\rangle\langle b_{k}|\bigr)^{\otimes n},
\end{equation}
which imprints a mode-dependent relative phase $e^{\imath nb_{k}\lambda}$
on each component $|b_{k}\rangle^{\otimes n}$, yielding 
\begin{equation}
\ensuremath{e^{\imath n\lambda\hat{\mathcal{M}}}|\mathrm{\mathcal{M}}_{d}^{(n)}\rangle=\frac{1}{\sqrt{d}}\sum_{k=0}^{d-1}e^{\imath nb_{k}\lambda}|b_{k}\rangle^{\otimes n}}.
\end{equation}
This ensures a phase sensitivity of the meter state scales linearly
with the number of subsystems $n$. The operator $e^{\imath n\lambda\hat{\mathcal{M}}}$
reduces to the familiar single-qubit phase gate in case $d=2,\,\,n=1$,
thus providing a natural generalization to multipartite, high-dimensional
entangled meter states (see Appendix \ref{sec:Derivation-of-the} for more details). By acting on $|\mathrm{\mathcal{M}}_{d}^{(n)}\rangle$
using the channel operator we get
\begin{align}
\hat{\mathds{K}}_{m_{f},m_{i}}^{(j)}|\mathrm{\mathcal{M}}_{d}^{(n)}\rangle= & \frac{e^{\imath j\Theta}}{\sqrt{d}}\sum_{k=0}^{d-1}e^{\imath jnb_{k}\lambda}d_{m_{f},m_{i}}^{(j)}{\textstyle {\scriptstyle {\displaystyle (\beta_{k})}}}|b_{k}\rangle^{\otimes n},\label{eq:k=000020on=000020m}
\end{align}
where $\beta_{k}=\Theta-nb_{k}\lambda$. Thus, the postselection probability
gives
\begin{align}
\mathcal{P}_{m_{f},m_{i}}^{(j)}= & \frac{1}{d}\sum_{k=0}^{d-1}\left[d_{m_{f},m_{i}}^{(j)}{\textstyle {\scriptstyle {\displaystyle (\beta_{k})}}}\right]^{2},\label{eq:postselection=000020probability}
\end{align}
here the squared $d$-matrix elements quantify the transition probabilities
between the states $|j,m_{i}\rangle$ and $|j,m_{f}\rangle$ induced
by a rotation through the angle $\beta_{k}$. Averaging these elements
over $d$ terms reflects a uniform sampling of the rotation parameter
$\beta_{k}$, ensuring an unbiased characterization. For the Wigner
d-matrix $\ensuremath{\sum_{m_{f}}d_{m_{f},m_{i}}^{(j)}(\beta_{k})d_{m_{f},m_{j}}^{(j)}(\beta_{k})=\delta_{m_{i},m_{j}}}$,
and hence $\ensuremath{\sum_{m_{f}}\bigl[d_{m_{f},m_{i}}^{(j)}(\beta_{k})\bigr]^{2}=1}$,
which confirms that $\ensuremath{\sum_{m_{f}}\mathcal{P}_{m_{f},m_{i}}^{(j)}=1}$.

Taking the derivative of Eq. (\ref{eq:k=000020on=000020m}) with respect
to the parameter $\lambda$ gives
\begin{equation}
\partial_{\lambda}\hat{\mathds{K}}_{m_{f},m_{i}}^{(j)}|\mathrm{\mathcal{M}}_{d}^{(n)}\rangle=\frac{ne^{\imath j\Theta}}{\sqrt{d}}\sum_{k=0}^{d-1}b_{k}e^{\imath jnb_{k}\lambda}A_{k}|b_{k}\rangle^{\otimes n},
\end{equation}
where $A_{k}:=\imath jd_{m_{f},m_{i}}^{(j)}(\beta_{k})-\frac{d}{d\beta}d_{m_{f},m_{i}}^{(j)}(\beta)\bigr|_{\beta=\beta_{k}}$.
The total change in the channel operator captured by $\mathcal{Q}^{T}$
can subsequently be expressed as
\begin{align}
\ensuremath{\ensuremath{\mathcal{Q}^{T}=}} & \frac{n^{2}}{d}\sum_{k=0}^{d-1}\sum_{k^{\prime}=0}^{d-1}b_{k}u_{k^{\prime}}e^{-\imath jnu_{k^{\prime}}\lambda}e^{\imath jnb_{k}\lambda}\nonumber \\
 & \times A_{k}A_{k^{\prime}}^{*}\langle b_{k^{\prime}}|^{\otimes n}|b_{k}\rangle^{\otimes n}\nonumber \\
= & \ensuremath{\frac{n^{2}}{d}\sum_{k=0}^{d-1}b_{k}^{2}\bigl|A_{k}\bigr|^{2}},\label{eq:TOTAL}
\end{align}
where $\langle b_{k^{\prime}}|^{\otimes n}|b_{k}\rangle^{\otimes n}=\delta_{k,k^{\prime}}$.
Equivalently, the parallel term takes the form
\begin{align}
\mathcal{Q}^{\parallel}=\frac{n^{2}}{d^{2}} & \bigl|\sum_{k=0}^{d-1}b_{k}d_{m_{f},m_{i}}^{(j)}(\beta_{k}).A_{k}\bigr|^{2}.\label{eq:PARRALLEL=000020TERM}
\end{align}

\section{Results and discussion}

\subsection{Meter Operator Design}

Here, we analyze two representative classes of meter states that critically
shapes and maximizes the QFI. 

The first case corresponds to the generator $\hat{\mathcal{M}}=\hat{J}_{t}^{(\mathcal{M})}+\hat{J}_{z}^{(\mathcal{M})}$, which leads to  the computational basis for
the $n$-particle system $\ensuremath{\mathcal{B}=\{|k_{i}\rangle^{\otimes n}}\},$
where $k_{i}\in\{0,1,\dots,d-1\}$, and $|k\rangle$ denoting the
Fock state with $k$ bosons in the meter mode. The meter state represents a multimode boson state that that can be defined as $|\mathrm{\mathcal{M}}_{d}^{(n)}\rangle=\frac{1}{\sqrt{d}}\sum_{k=0}^{d-1}|k\rangle^{\otimes n}$.
The corresponding meter operator is given by
\begin{equation}
\hat{\mathcal{O}}_{\lambda}^{P}=e^{\imath n\lambda\hat{\mathcal{M}}}=\sum_{k=0}^{d-1}e^{\imath nk\lambda}(|k\rangle\langle k|\bigr)^{\otimes n}.
\end{equation}
We denote this meter operator by $\hat{\mathcal{O}}_{\lambda}^{P}$
to emphasize its role as the generator of the Pancharatnam phase
\begin{equation}
\textrm{Im\,ln}\langle\hat{\mathcal{O}}_{\lambda}^{P}\rangle=\textrm{Im\,Ln}(\sum_{k=0}^{d-1}e^{\imath nk\lambda})=n\frac{(d-1)\lambda}{2}.
\end{equation}
This operator enables efficient $\Theta$-optimization of $\mathcal{Q}^{\parallel}$
since
\begin{align}
\left\langle \!\right.(\hat{\mathcal{O}}_{\lambda}^{P})^{\dagger}\partial_{\lambda}\hat{\mathcal{O}}_{\lambda}^{P}\left.\!\right\rangle = & \ensuremath{\frac{1}{d}\sum_{k=0}^{d-1}\langle k|^{\otimes n}(\hat{\mathcal{O}}_{\lambda}^{P})^{\dagger}\partial_{\lambda}\hat{\mathcal{O}}_{\lambda}^{P}|k\rangle^{\otimes n}}\nonumber \\
= & \ensuremath{\ensuremath{\imath n\frac{1}{d}\sum_{k=0}^{d-1}k}=\ensuremath{in\frac{d-1}{2}\neq0.}}
\end{align}

The second case corresponds to the generator $\hat{\mathcal{M}}=\hat{J}_{z}^{(\mathcal{M})}$, which produces the angular-momentum-basis, $|b_{k}\rangle=|\tilde{j},m_{k}\rangle$,
where symbol $\tilde{j}$ is used to distinguish it from the $j$
of the postselected system. Here $m_{k}=k-\tilde{j}$ runs over all
integer or half-integer values from $-\tilde{j}$ to $\tilde{j}$,
and $\tilde{j}=\frac{d-1}{2}$. This yields
\begin{equation}
|\mathrm{\mathcal{M}}_{d}^{(n)}\rangle=\frac{1}{\sqrt{d}}\sum_{k=0}^{d-1}|\tilde{j},k-\tilde{j}\rangle^{\otimes n}=\frac{1}{\sqrt{d}}\sum_{m=-\tilde{j}}^{\tilde{j}}|\tilde{j},m\rangle^{\otimes n},
\end{equation}
with the angular momentum basis being simply a reindexed computational
basis. The meter operator in this case takes the form
\begin{equation}
\hat{\mathcal{O}}_{\lambda}^{(0)}=e^{\imath n\lambda\hat{\mathcal{M}}}=\sum_{m=-\tilde{j}}^{\tilde{j}}e^{\imath n m\lambda}(|\tilde{j},m\rangle\langle\tilde{j},m|)^{\otimes n}
\end{equation}

We label this operator as $\hat{\mathcal{O}}_{\lambda}^{(0)}$ to
indicate the absence of the Pancharatnam phase
\begin{equation}
\textrm{Im\,ln}\langle\hat{\mathcal{O}}_{\lambda}^{(0)}\rangle=\textrm{Im\,Ln}(\sum_{m=-\tilde{j}}^{\tilde{j}}e^{\imath nm\lambda})=0.
\end{equation}
The operator $\hat{\mathcal{O}}_{\lambda}^{(0)}$ obstructs the $\Theta$-optimization
of $\mathcal{Q}^{\parallel}{\scriptstyle (\Theta)}$ because
\begin{align}
\left\langle \!\right.(\hat{\mathcal{O}}_{\lambda}^{(0)})^{\dagger}\partial_{\lambda}\hat{\mathcal{O}}_{\lambda}^{(0)}\left.\!\right\rangle  & =\ensuremath{\frac{1}{d}\sum_{m=-\tilde{j}}^{\tilde{j}}\langle\tilde{j},m|^{\otimes n}(\hat{\mathcal{O}}_{\lambda}^{(0)})^{\dagger}\partial_{\lambda}\hat{\mathcal{O}}_{\lambda}^{(0)}|\tilde{j},m\rangle^{\otimes n}}\nonumber \\
 & =\ensuremath{\frac{\imath n}{d}\sum_{m=-\tilde{j}}^{\tilde{j}}m}=0.
\end{align}
 As we shall see below, the QFI in the latter case is smaller than
in the former. Remarkably, however, the two cases are physically equivalent
with respect to the meter state, differing only by an overall phase
factor. With the generators $\hat{\mathcal{M}}=\hat{J}_{t}^{(\mathcal{M})}+\hat{J}_{z}^{(\mathcal{M})}$
and $\hat{\mathcal{M}}=\hat{J}_{z}^{(\mathcal{M})}$
of $\hat{\mathcal{O}}_{\lambda}^{P}$ and $\hat{\mathcal{O}}_{\lambda}^{(0)}$
respectively, we write
\begin{equation}
\hat{\mathcal{O}}_{\lambda}^{P}=e^{\imath n\lambda(\hat{J}_{t}^{(\mathcal{M})}+\hat{J}_{z}^{(\mathcal{M})})}=e^{\imath n\lambda\hat{J}_{t}^{(\mathcal{M})}}\hat{\mathcal{O}}_{\lambda}^{(0)}.
\end{equation}
For a fixed dimension of the meter subspace, $\hat{J}_{t}^{(\mathcal{M})}$
acts as multiplication by a scalar $\tilde{j}$, contributing a global
phase factor $e^{\imath n\tilde{j}\lambda}$, independent of the basis states
$|k\rangle$ or $|\tilde{j},m\rangle$. For example, when $d=3$ ($\tilde{j}=1$):
\begin{equation}
\hat{\mathcal{O}}_{\lambda}^{P}=\left(|0\rangle\langle0|+e^{\imath\lambda}|1\rangle\langle1|+e^{2\imath\lambda}|2\rangle\langle2|\right)^{\otimes n},
\end{equation}
and
\begin{align}
\hat{O}_{\lambda}^{(0)} & =\left(e^{-\imath\lambda}|1,-1\rangle\langle1,-1|+|1,0\rangle\langle1,0|+e^{\imath\lambda}|1,1\rangle\langle1,1|\right)^{\otimes n}\nonumber \\
 & \equiv e^{-\imath n\lambda}\left(|0\rangle\langle0|+e^{\imath\lambda}|1\rangle\langle1|+e^{2\imath\lambda}|2\rangle\langle2|\right)^{\otimes n}\nonumber \\
 & =e^{-\imath n\lambda}\hat{\mathcal{O}}_{\lambda}^{P},
\end{align}
where $e^{-\imath n\lambda}$ compensates for the index shift $k=m+1$.
Considering the structure of the interaction unitary operator in Eq.
(\ref{eq:Interaction=000020unitary=000020operator}), the operator
$\hat{\mathcal{O}}_{\lambda}^{P}$ corresponds to an experimental
scenario where a single system  mode interacts with a single meter
mode, whereas $\hat{O}_{\lambda}^{(0)}$ describes an interaction
involving the two meter modes. Notably, both operators generate the same
relative phase in the meter state.

\subsection{The Pancharatnam Phase Shift for Arbitrary
$j$}

In this section, we verify the previously stated result that the postselected system's phase shift is  characterized by the identical Pancharatnam phase, independent of the quantum number $j$ and irrespective of the specific choices of pre- and postselection states.

Consider pre- and postselecting the highest and lowest weight states denoted by  $|j,m_{i}\rangle=|j,j\rangle$,
and $|j,m_{f}\rangle=|j,-j\rangle$, respectively, wherein the relevant Wigner d-matrix element takes the form $d_{-j,j}^{(j)}{\scriptstyle {\displaystyle \bigl(\beta_{k}\bigr)}}=\bigl[\sin(\frac{\beta_{k}}{2})\bigr]^{2j}$, with $\beta_{k}=\Theta-nk\lambda.$.
The expressions given in (\ref{eq:postselection=000020probability}),
(\ref{eq:TOTAL}), and (\ref{eq:PARRALLEL=000020TERM}) can then be, respectively,
rewritten as
\begin{align}
\mathcal{P}_{-j,j}^{(j)}= & \frac{1}{d}\sum_{k=0}^{d-1}\bigl[\sin(\frac{\beta_{k}}{2})\bigr]^{4j},\label{eq:27=000020postselection}\\
\mathcal{Q}^{T}= & \frac{n^{2}j^{2}}{d}\sum_{k=0}^{d-1}b_{k}^{2}\sin^{4j-2}(\frac{\beta_{k}}{2})\label{eq:total=0000201},\\
\mathcal{Q}^{\parallel}= & \frac{n^{2}j^{2}}{d^{2}}\bigl|\sum_{k=0}^{d-1}b_{k}\sin^{4j-1}(\frac{\beta_{k}}{2})e^{\imath\frac{nb_{k}\lambda}{2}}\bigr|^{2}.\label{eq:q=000020parallel}
\end{align}
representing the postselection probability, total and parallel changes.

When the transition between pre- and postselected states is by a single
quantum with $m_{i}=j$ and $m_{f}=j-1$ , the corresponding Wigner
d-matrix element takes the well-known form
\begin{equation}
d_{j-1,j}^{(j)}{\displaystyle \bigl(\beta\bigr)}=-\sqrt{2j}\sin\frac{\beta_{k}}{2}(\cos\frac{\beta_{k}}{2})^{2j-1}.\label{eq:jtoj-1}
\end{equation}
Using this expression, the postselection probability can be written
as \begin{align}
\mathcal{P}_{j-1,j}^{(j)}= & \frac{2j}{d}\sum_{k=0}^{d-1}\bigl[\sin\frac{\beta_{k}}{2}(\cos\frac{\beta_{k}}{2})^{2j-1}\bigr]^{2}\label{eq:postselection2}
\end{align}
Notably, for $j=1/2,1$, exact closed-form expressions for the postselection probability are obtained
\begin{align}
\mathcal{P}_{-\frac{1}{2},\frac{1}{2}}^{(\frac{1}{2})} & =\ensuremath{\frac{1}{2}\bigl[1-\frac{\sin\bigl(\frac{dn\lambda}{2}\bigr)}{d\sin\bigl(\frac{n\lambda}{2}\bigr)}\cos\bigl(\Theta-\frac{(d-1)n\lambda}{2}\bigr)\bigr],}\\
\mathcal{P}_{0,1}^{(1)} & =\ensuremath{\frac{1}{4}\bigl[1-\frac{\sin\bigl(dn\lambda\bigr)}{d\sin\bigl(n\lambda\bigr)}\cos2\bigl(\Theta-\frac{(d-1)n\lambda}{2}\bigr)\bigr].}
\end{align}
For higher $j$ values, the postselection probabilities generalize
to sums involving harmonic terms $S_{\sigma}$ 
\begin{align}
\ensuremath{\mathcal{P}_{\frac{1}{2},\frac{3}{2}}^{(\frac{3}{2})}} & =\frac{3}{16}(1+\frac{1}{2}S_{1}-S_{2}-\frac{1}{2}S_{3}),\\
\mathcal{P}_{1,2}^{(2)}& =\frac{1}{8}(\frac{5}{4}+S_{1}-S_{2}-S_{3}-\frac{1}{4}S_{4}),
\end{align}
where each $S_{\sigma}$ encodes multi-path interference effects through
\begin{equation}
S_{\sigma}=\ensuremath{\frac{\sin\bigl(\sigma\frac{dn\lambda}{2}\bigr)}{d\sin\bigl(\sigma\frac{n\lambda}{2}\bigr)}\cos \sigma\bigl(\Theta-\frac{(d-1)n\lambda}{2}\bigr)}
\end{equation}
where $\sigma=1,2,3,\ldots.$

\begin{figure}[t]
\begin{centering}
\includegraphics[width=1\columnwidth]{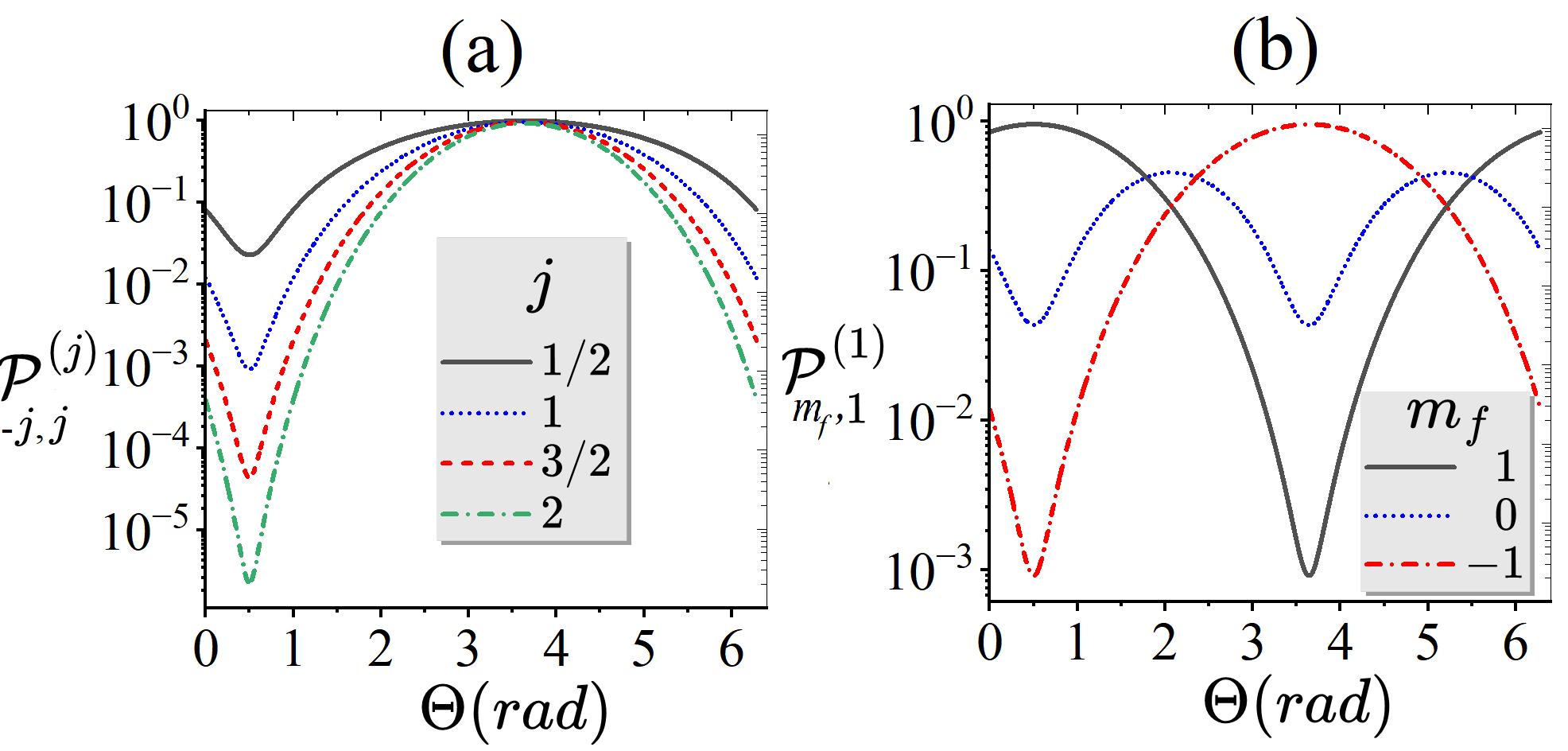}
\par\end{centering}
\caption{Postselection probability as a function of the postselection parameter
$\Theta$, for $\lambda=10^{-3}rad$, $d=21,n=50$. This produces a Pancharatnam phase shift $\textrm{Im}\ln\langle\hat{\mathcal{O}}_{\lambda}^{P}\rangle=\frac{(d-1)n\lambda}{2}=0.5$. (a) For different values
of $j$ with preselected state $m_{i}=j,$  and postselected state $m_{f}=-j$. (b)  For $j=1$, $m_{i}=j$ and different values of $m_{f}$.}\label{fig:0}
\end{figure}

Figure \ref{fig:0} presents the postselection probability $\mathcal{P}_{m_{f},m_{i}}^{(j)}$
as a function of the parameter $\Theta$ for various quantum numbers
j, alongside results for fixed $j=1$ with varying postselected states
$m_{f}$. While the visibility and oscillation frequency of the interference
fringes depend on these quantum numbers, the Pancharatnam phase shift $\textrm{Im}\ln\langle\hat{\mathcal{O}}_{\lambda}^{P}\rangle$
emerges as the fundamental invariant quantity governing the postselection probability of the compression
channels.

\subsection{Complete Nullification of $\mathcal{Q}^{\parallel}$ for Arbitrary
$j$}\label{Complete Nullification}

As introduced in Section \ref{Controlling Parallel Evolution}, by post-selecting on system characterized
by the Pancharatnam phase, one gains direct control over the
Parallel contribution to QFI in the quantum channel, enabling its
complete suppression and rendering the compression channel lossless.

To demonstrate this quantitatively for a general quantum number $j$,
we consider the expressions derived from the Wigner d-matrix formalism.
The total $\mathcal{Q}^{T}$ and parallel $\mathcal{Q}^{\parallel}$
components can be expressed as
\begin{align}
\mathcal{Q}^{T}= &2 \frac{n^{2}j}{d}\sum_{k=0}^{d-1}b_{k}^{2}|A_{k}|^{2}\label{eq:total2-1}\\
\mathcal{Q}^{\parallel}= & 4\frac{n^{2}j^{2}}{d^{2}}\bigl|\sum_{k=0}^{d-1}b_{k}\sin\frac{\beta_{k}}{2}(\cos\frac{\beta_{k}}{2})^{2j-1}A_{k}\bigr|^{2}\label{eq:q=000020parallel2-1}
\end{align}
where the complex amplitudes $A_{k}$ are given by \begin{align*}
A_{k}= & \imath j(\sin\frac{\beta_{k}}{2}\cos^{2j-1}\frac{\beta_{k}}{2})\\
 & -\frac{1}{2}\cos^{2j-2}\frac{\beta_{k}}{2}[\cos^{2}\frac{\beta_{k}}{2}-(2j-1)\sin^{2}\frac{\beta_{k}}{2}].
\end{align*}

For the operator $\hat{\mathcal{O}}_{\lambda}^{P}$, in the simplest nontrivial case $d=2$, the parallel component simplifies
to
\begin{align}
\mathcal{Q}^{\parallel}= & \ensuremath{\frac{n^{2}j^{2}}{4}\sin^{2}(\frac{\Theta-n\lambda}{2})\cos^{8j-6}(\frac{\Theta-n\lambda}{2})}\nonumber \\
 & \times[\cos^{2}(\frac{\Theta-n\lambda}{2})+(2j-1)^{2}\sin^{2}(\frac{\Theta-n\lambda}{2})].
 \label{parrallel d=2}
\end{align}
By choosing $\Theta=n\lambda$, this term is identically null for
all $j$, demonstrating that the parallel contribution to the QFI can
be completely suppressed through this parameter tuning. 

For the operator $\hat{\mathcal{O}}_{\lambda}^{(0)}$, the parallel
component of the QFI simplifies to 
$\mathcal{Q}^{\parallel}=\frac{n^{2}}{16}\sin^{2}\left(\frac{n\lambda}{2}\right)\neq0$
for $j=\frac{1}{2}$, and 
$
\mathcal{Q}^{\parallel}=\frac{n^{2}}{64}\sin^{2}(n\lambda)\neq0
$
for $j=1$. Analytical expressions become intractable for higher $j$,
yet in the small-$\lambda$ regime, one obtains the asymptotic scaling, for instance, when $j=3/2$,
$
\mathcal{Q}^{\parallel}\approx\frac{9n^{4}}{256}\lambda^{2}.
$
This behavior reveals a fundamental limitation on loss mitigation
achievable through compression channels.

\begin{figure*}[t]
\centering{}\includegraphics[width=1\textwidth]{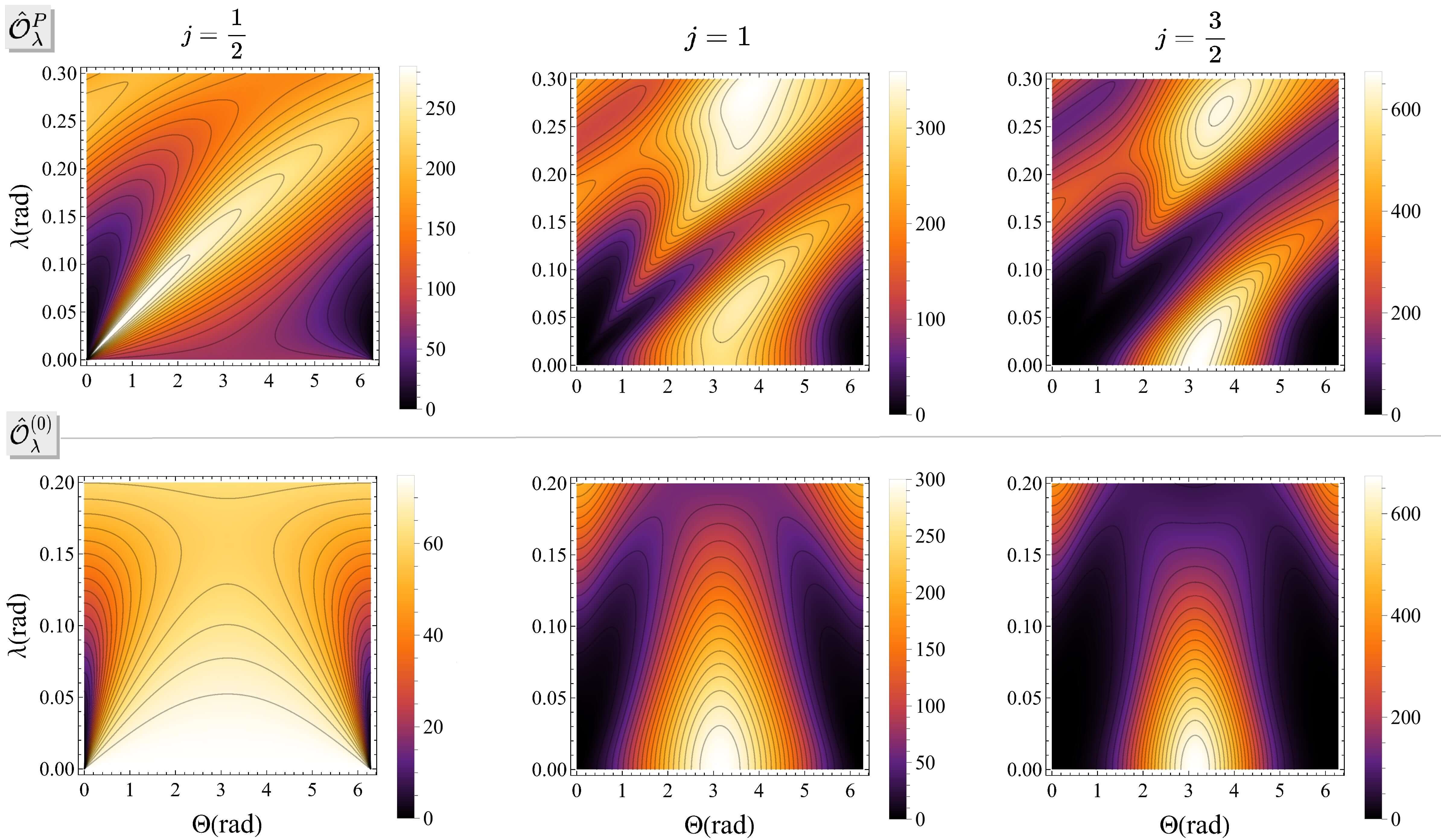}\caption{Contour plots of the  QFI per trial $\mathcal{T}{\scriptstyle (\lambda,\Theta)} $
depicted as functions of the coupling and postselection parameters
$\lambda$ and $\Theta$, for quantum numbers $j=1/2$, $1$, and $3/2$, $n=1$, $ d=30$.
The top row corresponds to calculations using the meter operator $\hat{O}_{\lambda}^{P},$
while the bottom row displays results obtained with $\hat{O}_{\lambda}^{(0)}$.
In each panel, the color scale encodes the $\mathcal{T}{\scriptstyle (\lambda,\Theta)}$ magnitude,
with lighter shades indicating higher sensitivity. Further discussion
and detailed physical interpretation are provided in the main text.}\label{fig:1}
\end{figure*}

It is important to highlight that Eq.~(\ref{parrallel d=2}) encompasses parameter
regimes where the parallel term vanishes, specifically for cases with
$j\geq\frac{3}{2}$ and $\Theta=\pi+n\lambda$. This behavior aligns
with a high postselection probability regime, indicating no compression
of metrological information, as evident from the postselection probability
profile Eq. (\ref{eq:postselection2}). In such scenarios,  contingent on the interaction strength $\lambda$ and meter dimensionality $d$, we emphasize the possibility of exploring
alternative postselection schemes on single quantum transitions---such
as $d_{0,1}^{j}$ for integer $j$ and $d_{-\frac{1}{2},\frac{1}{2}}^{j}$
for half-integer $j$---to realize a lossless compression channel.
The forthcoming analysis illustrates the core concept using Eq.~\eqref{parrallel d=2},
but for larger values of $j$, the aforementioned alternatives remain
valid, adhering to the same methodological framework and logic presented
here.

\subsection{QFI Yield: $d_{j-1,j}^{(j)}$ versus  $d_{-j,j}^{(j)}$}

It is well established that compressing metrological information into a small subset of postselection outcomes requires the success events to be intrinsically rare. However, this condition is necessary but not sufficient. Crucially, the postselection must be performed on a single quantum transition to realize a lossless quantum compression channel.

Figure \ref{fig:1} presents contour plots of the total QFI per trial $\mathcal{T}{\scriptstyle (\lambda,\Theta)}=\mathcal{P}_{-j,j}^{(j)}\mathcal{I}^{\perp}$
as a function of the coupling parameter $\lambda$ and $\Theta$,
for various quantum numbers $j$. The results correspond to $d=30$,
obtained directly from Eqs. (\ref{eq:27=000020postselection}), (\ref{eq:total=0000201}),
and (\ref{eq:q=000020parallel}). 

For $j=1/2$, the operator $\hat{\mathcal{O}}_{\lambda}^{P}$
displays a pronounced diagonal ridge even at small values of $\lambda$,
indicating a significant enhancement in quantum sensitivity, $\mathcal{T} {\scriptstyle (\lambda,\Theta)}\gg 1$,
along correlated values of $\lambda$ and $\Theta$. In contrast,
the operator $\hat{\mathcal{O}}_{\lambda}^{(0)}$ exhibits a broad, symmetric
maximum centered at small $\lambda$ and intermediate $\Theta$, decreasing
rapidly outside this region. Overall, the quantum advantage in sensitivity
achievable with $\hat{\mathcal{O}}_{\lambda}^{(0)}$ is notably limited
compared to that of $\hat{\mathcal{O}}_{\lambda}^{P}$.

For higher values\textbf{ $j=1,3/2$},\textbf{ }the landscape exhibits
multiple local maxima and complex contour structures, with prominent
high-$\mathcal{T} {\scriptstyle (\lambda,\Theta)}$ regions occurring at intermediate parameter values.
This behavior highlights the suboptimality of the small $(\Theta,\lambda)$
regime as $j$ increases, where $\mathcal{T}{\scriptstyle (\lambda,\Theta)}\ll 1$. 
For smaller $\lambda$, $\mathcal{T}{\scriptstyle (\lambda,\Theta)}$ reaches its highest amplitudes
near $\Theta\approx\pi$, corresponding to a high postselection probability (no compression).
As the quantum number $j$ increases, these peaks become increasingly
localized around $\Theta\approx\pi$ and remain constrained to the small $\lambda$ regime. 

\begin{figure} 
\begin{centering}
\includegraphics[width=1\columnwidth]{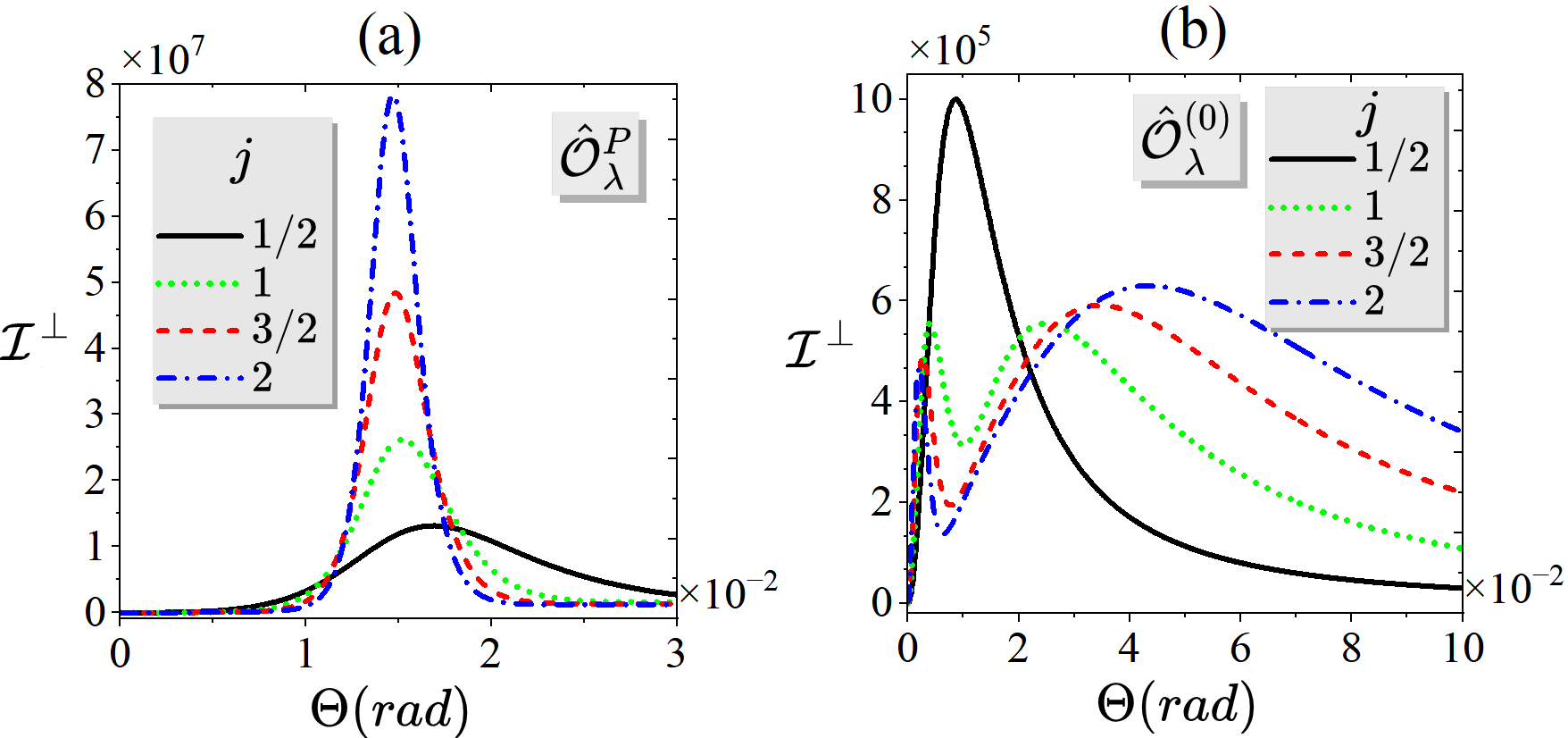}
\par\end{centering}
\caption{The postselected QFI  $\mathcal{I}^{\perp}$ as a function of $\Theta$ for $n\lambda=10^{-3}$ and $d=30$. (a) QFI exhibits
a linear   growth as a function of $j$.  Notably, the QFI peak shifts towards the Pancharatnam phase
as $j$ increases. (b) Reference protocol $\hat{\mathcal{O}}_{\lambda}^{(0)}$  shows qualitative differences in scaling and suppression of QFI as $j$ increases. }\label{fig:2}
\end{figure}
\begin{figure*}[t]
\includegraphics[width=0.9\textwidth]{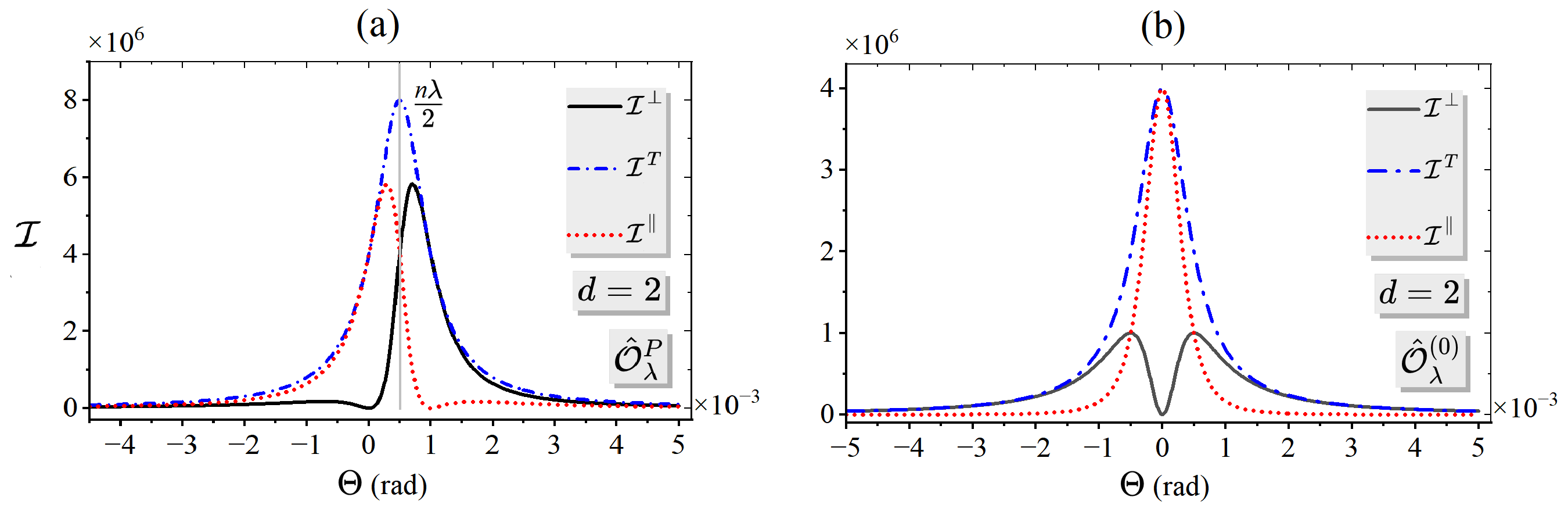}
\caption{Postselected QFI $\mathcal{I}^{\perp}$, $\mathcal{I}^{T}$,
and $\mathcal{I}^{\parallel}$ as a function of $\Theta$
for $n\lambda=10^{-3}$,   $d=2$ and arbitrary $j$. (a) For the meter operator $\hat{\mathcal{O}}_{\lambda}^{P}$,
the characteristic trade-off between $\mathcal{I}^{\perp}$
and $\mathcal{I}^{\parallel}$ occurs
in the vicinity of the Pancharatnam phase $\frac{n\lambda}{2}$. (b)
For the meter operator $\hat{\mathcal{O}}_{\lambda}^{(0)}$, both
$\mathcal{I}^{\parallel}$ and $\mathcal{I}^{T}$
exhibit cooperative behavior, whereas the postselected QFI $\mathcal{I}^{\perp}$
suffers an approximately 83\% reduction at $\Theta_{\perp}=\pm\frac{n\lambda}{2}$,
indicating suboptimal performance of operator $\hat{\mathcal{O}}_{\lambda}^{(0)}$
in comparison with $\hat{\mathcal{O}}_{\lambda}^{P}$.}\label{Figure4}
\end{figure*}

This behavior arises because the postselection probability decreases exponentially with increasing quantum number  
$j$, as illustrated in Fig. \ref{fig:0}(a). Such decay can be well approximated by the expression:
\begin{align}
\mathcal{P}_{-j,j}^{(j)}\approx\bigl[\mathcal{P}_{-\frac{1}{2},\frac{1}{2}}^{(\frac{1}{2})}\bigr]^{2j},
\end{align}
whereas, the QFI grows linearly with $j$, as shown in Fig. \ref{fig:2}(a). Consequently, the total QFI
per trial $\mathcal{T} {\scriptstyle (\lambda,\Theta)}$ decreases rapidly with increasing $j$ in the weak-coupling
regime.

To achieve optimal quantum compression for higher \(j\) values, we postselect on  quantum states that induce a single quantum transition given by Eq. (\ref{eq:jtoj-1}). The postselection probability for this case is given by Eq.~(\ref{eq:postselection2}) where $\mathcal{P}_{j-1,j}^{(j)}\sim j\mathcal{P}_{-\frac{1}{2},\frac{1}{2}}^{(\frac{1}{2})}$. This procedure restores the metrological information in $\mathcal{T} {\scriptstyle (\lambda,\Theta)}$ to \(j\) times that of the \(j = \tfrac{1}{2}\) case, Eq.~(\ref{eq:total2-1}), thereby generalizing the optimal compression to arbitrary \(j\).

It is straightforward to verify whether the compression channel associated
with the operator $\hat{\mathcal{O}}_{\lambda}^{P}$ at $\Theta=\Theta_{\parallel}$
is lossless. The QFI per trial is given by 
\[
\mathcal{T}{\scriptstyle (\lambda,\Theta_{\parallel})}=\mathcal{P}_{j-1,j}^{(j)}\mathcal{I}^{\perp}=4Q^{T},
\]
which exactly matches the standard QFI for pure states without postselection
\[
\mathcal{I}_{S}=
\sum_{m_{f}}\frac{1}{\mathcal{P}_{m_{f},j}^{(j)}}(\frac{\partial\mathcal{P}_{m_{f},j}^{(j)}}{\partial\lambda})^{2},
\]
assuming all quantum systems are initially prepared in the state $m_{i}=j$. For example, when $d=2,\,n=1$, it can be shown that the QFI per trial reduces to $\mathcal{T}{\scriptstyle (\lambda,\Theta_{\parallel})}=\mathcal{I}_{S}=j$, taking into careful consideration the final remark presented in Section \ref{Complete Nullification}.

\subsection{Poselected QFI $\mathcal{I}^{\perp}$ Analysis for a Qubit Meter}

For $d=2,$ the operator $\hat{\mathcal{O}}_{\lambda}^{P}$ can
be defined as
\begin{equation}
\hat{\mathcal{O}}_{\lambda}^{P}=|0\rangle\langle0|^{\otimes n}+e^{\imath n\lambda}|1\rangle\langle1|^{\otimes n},
\end{equation}
which imparts relative phase $n\lambda$ to the meter states. Given
the initial state of the meter $|\mathrm{\mathcal{M}}_{2}^{(n)}\rangle=\frac{1}{\sqrt{2}}(|0\rangle^{\otimes n}+|1\rangle^{\otimes n})$,
the Pancharatnam phase is expressed as $\textrm{Im\,ln}\left\langle \!\right.\hat{\mathcal{O}}_{\lambda}^{P}\left.\!\right\rangle =n\frac{\lambda}{2}$.

From Eq. (\ref{eq:total2-1}), the total contribution for QFI $\mathcal{Q}^{T}$ for different $j$ values 
are given by \begin{equation}
\begin{aligned} & j=\frac{1}{2}:\frac{n^{2}}{8},\\
 & j=1:\frac{n^{2}}{4},\\
 & j=\frac{3}{2}:\frac{3n^{2}}{2}\cos^{2}\frac{\Theta-n\lambda}{2}(1-\frac{3}{4}\cos^{2}\frac{\Theta-n\lambda}{2}).
\end{aligned}
\end{equation}

 The same holds for the term corresponding to parallel evolution $\mathcal{Q}^{\parallel}$. From Eq. (\ref{parrallel d=2}), we find $\mathcal{Q}^{\parallel}$
\begin{equation}
\begin{aligned} & j=\frac{1}{2}:\frac{n^{2}}{16}\sin^{2}\frac{\Theta-n\lambda}{2},\\
 & j=1:\frac{n^{2}}{4}\sin^{2}\frac{\Theta-n\lambda}{2}\cos^{2}\frac{\Theta-n\lambda}{2},\\
 & j=\frac{3}{2}:\frac{9n^{2}}{16}\sin^{2}\frac{\Theta-n\lambda}{2}\cos^{6}\frac{\Theta-n\lambda}{2}\\
 & \qquad\qquad\times(\cos^{2}\frac{\Theta-n\lambda}{2}+4\sin^{2}\frac{\Theta-n\lambda}{2}).
\end{aligned}
\end{equation}

\begin{figure*}[t]
\includegraphics[width=1\textwidth]{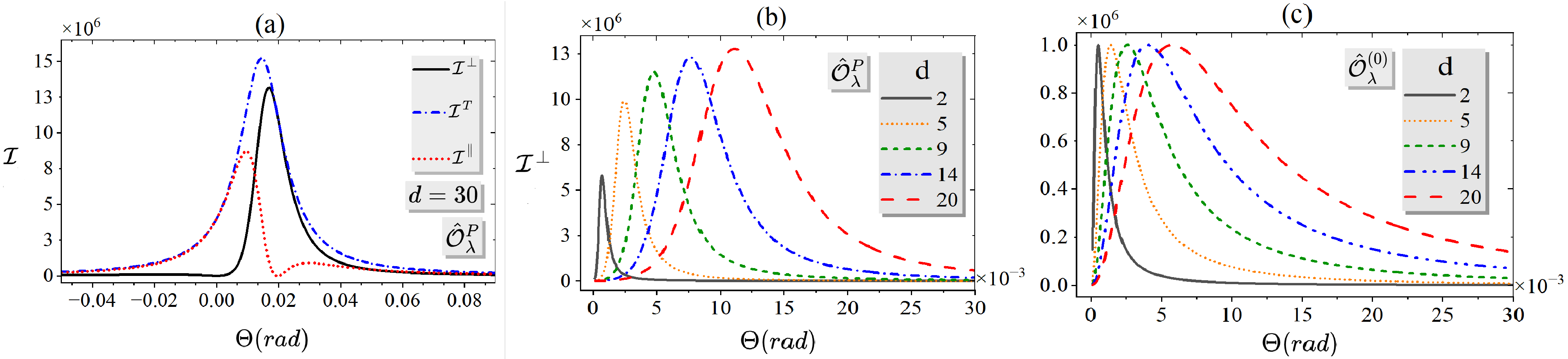}

\caption{(a) Variation of the postselected QFI $\mathcal{I}^{\perp}$,
$\mathcal{I}^{T}$, and $\mathcal{I}^{\parallel}$
as functions of $\Theta$ for $d=30$, $n\lambda=10^{-3}$ and arbitrary $j$, illustrating
how the operator $\hat{\mathcal{O}}_{\lambda}^{P}$ enables precise
tuning of the trade-off between orthogonal and parallel sensitivities.
(b) The  QFI $\mathcal{I}^{\perp}$
plotted against $\Theta$ for various meter dimensions $d$, demonstrating
a pronounced increase in QFI and the formation of sharply localized
peaks with rising $d$ when using $\hat{\mathcal{O}}_{\lambda}^{P}$.
These peaks signify optimal $\Theta$ at which maximal quantum sensitivity
is achieved as meter complexity grows. (c) In contrast, the QFI   $\mathcal{I}^{\perp}$
corresponding to $\hat{\mathcal{O}}_{\lambda}^{(0)}$ remains essentially
unchanged in its maximal numerical value as $d$ increases, reflecting
a diminished sensitivity to the meter dimension $d$. }\label{fig:4}
\end{figure*}

Accordingly, taking into account  Eqs. (\ref{eq:total2-1}), (\ref{parrallel d=2}), and the postselection proability $\mathcal{P}_{j-1,j}^{(j)}$ in Eq. (\ref{eq:postselection2}), we can find the postselected QFI $\mathcal{I}^{\perp}=\mathcal{I}^{T}-\mathcal{I}^{\parallel},$ given by the Eq. (\ref{postselected QFI}),
with
\begin{align}
\mathcal{I}^{T}=4\mathcal{Q}^{T}(\mathcal{P}_{j-1,j}^{(j)})^{-1},\quad
\mathcal{I}^{\parallel}=4\mathcal{Q}^{\parallel}(\mathcal{P}_{j-1,j}^{(j)})^{-2} .
\end{align}
Figure \ref{Figure4} shows $\ensuremath{\mathcal{I}^{\perp},\mathcal{I}^{T},\mathcal{I}^{\parallel},}$ as a function of $\Theta $ for arbitrary $j$.   $\mathcal{I}^{T}$ attains
its maximum for $$\Theta_{T}=\underset{\Theta}{\textrm{arg}\,\textrm{max}\,}\mathcal{I}^{T}{\scriptstyle ({\lambda},\Theta)}=n\frac{\lambda}{2}$$
(Pancharatnam phase), whereas the parallel component $\mathcal{I}^{\parallel}$
vanishes under the condition $$\Theta_{\parallel}=\underset{\Theta}{\textrm{arg}\,\textrm{min}\,}\mathcal{I}^{\parallel}{\scriptstyle (\lambda,\Theta)}=n\lambda.$$

Under the small-angle approximation for $\lambda$ ($n\lambda<1\,\text{rad}$), and regardless of the value of $j$, The postselection parameter $\Theta_{\textrm{\ensuremath{\perp}}}$
that maximizes $\mathcal{I}^{\perp}$
is 
\begin{equation}
\Theta_{\textrm{\ensuremath{\perp}}}=\underset{\Theta}{\textrm{arg}\,\textrm{max}\,}\mathcal{I}^{\perp}{\scriptstyle (\lambda,\Theta)}=\cos^{-1}\bigl[\cos^{2}\bigl(n\frac{\lambda}{2}\bigr)\bigr],
\end{equation}
a direct function of the Pancharatnam phase. Figure~\ref{Figure4}(a) illustrates
the trade-off around Pancharatnam phase between $\mathcal{I}^{\perp}$
and $\mathcal{I}^{\parallel}$ as
$\Theta$ varies. 

Figure~\ref{Figure4}(b) shows that the operator $\hat{\mathcal{O}}_{\lambda}^{(0)}=e^{-\imath n\lambda/2}|0\rangle\langle0|^{\otimes n}+e^{\imath n\lambda/2}|1\rangle\langle1|^{\otimes n},$
which introduces the same relative phase $n\lambda$ to the meter
state, leads to a cooperative behavior between $\mathcal{I}^{\parallel}$
and $\mathcal{I}^{T}$ which reduces
the postselected QFI by 83.33\%, rendering $\hat{\mathcal{O}}_{\lambda}^{(0)}$
suboptimal. QFI in this case is maximized for $\Theta_{\textrm{\ensuremath{\perp}}}=\underset{\Theta}{\textrm{arg}\,\textrm{max}\,}\mathcal{I}^{\perp}{\scriptstyle (\lambda,\Theta)}=\pm n\frac{\lambda}{2}$. 

The same information compression, characterized by identical postselection
probabilities, occurs when $\Theta=\Theta_{\textrm{\ensuremath{\parallel}}}=n\lambda$
and $\Theta=\Theta_{\textrm{\ensuremath{\perp}}}=n\frac{\lambda}{2}$
for the operators $\hat{\mathcal{O}}_{\lambda}^{P}$ and $\hat{\mathcal{O}}_{\lambda}^{(0)}$
, respectively. The metrological gain in sensitivity to $\lambda$  of operator $\hat{\mathcal{O}}_{\lambda}^{P}$ relative to $\hat{\mathcal{O}}_{\lambda}^{(0)}$ can be quantified as 
\begin{equation}
g=10\log_{10}\frac{\textrm{\textbf{Var}}(\lambda)\bigr|_{\hat{\mathcal{O}}_{\lambda}^{(0)}}}{\textrm{\textbf{Var}}(\lambda)\bigr|_{\hat{\mathcal{O}}_{\lambda}^{P}}}=10\log_{10}\frac{\mathcal{I}^{\perp}{\scriptstyle (\Theta_{\textrm{\ensuremath{\parallel}}})}\bigr|_{\hat{\mathcal{O}}_{\lambda}^{P}}}{\mathcal{I}^{\perp}{\scriptstyle (\Theta_{\textrm{\ensuremath{\perp}}})}\bigr|_{\hat{\mathcal{O}}_{\lambda}^{(0)}}}=6\,\textrm{dB}
\end{equation}
which is comparable to the metrological gain recently demonstrated
in LIGO using squeezed light \citep{Lough2021}.

\subsection{Qudit-enhanced Sensitivity: Boosting $\mathcal{I}^{\perp}$, $\mathcal{T}$, $\Delta\lambda$, and $SNR$}

For a qudit meter with $d>2$, the postselection probability is determined by Eq. (\ref{eq:postselection2}). 
According to Eq.~(\ref{eq:total2-1}), 
$\mathcal{Q}^{T}$ for different $j$ values is expressed as
\begin{equation}
\begin{aligned} & j=\frac{1}{2}:\frac{n^{2}}{24}(d-1)(2d-1),\\
 & j=1:\frac{n^{2}}{12}(d-1)(2d-1),\\
 & j=\frac{3}{2}:\ensuremath{\frac{3n^{2}}{d}\sum_{k=0}^{d-1}k^{2}(\cos^{2}\frac{\beta_{k}}{2}-\frac{3}{4}\cos^{4}\frac{\beta_{k}}{2})}\\
 & \qquad\qquad\approx\frac{3n^{2}}{24}(d-1)(2d-1)~~~\textrm{for\,small \ensuremath{\beta_{k}}}.
\end{aligned}
\end{equation}

The analytical solution for the parallel component $Q^{\parallel}$
remains challenging to obtain for arbitrary values of $j$. However,
as discussed above, the explicit expression for $\Theta_{\parallel}$
can be derived for the specific case of $j=\tfrac{1}{2}$, and the
result is expected to hold for higher $j$ values. The accuracy and applicability
of the derived $\Theta_{\parallel}$ for larger $j$ can subsequently
be validated through numerical calculations.

\begin{figure*}[t]
\includegraphics[width=0.99\textwidth]{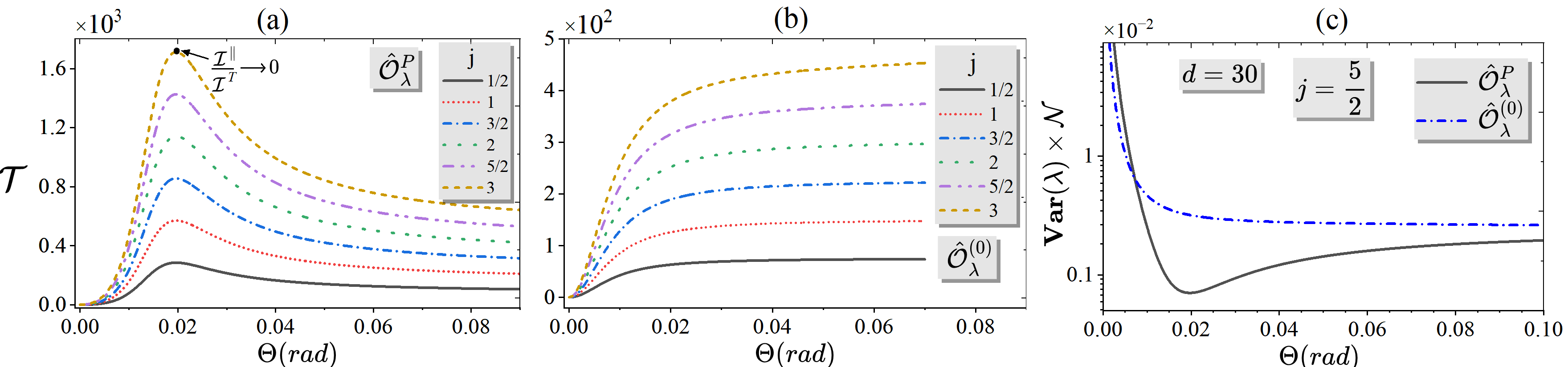}

\caption{Performance of $\hat{\mathcal{O}}_{\lambda}^{P}$ and $\hat{\mathcal{O}}_{\lambda}^{(0)}$
as a function of the postselection parameter $\Theta$, shown for
various values $j$, $d=30$, $\lambda=10^{-3}$ and $n=1$. (a) The estimator $\hat{\mathcal{O}}_{\lambda}^{P}$
realizes a pronounced enhancement in the QFI per trial, $\mathcal{T}$,
peaking at the optimal value $\Theta_{\parallel}$, where the parallel
contribution $\mathcal{I}^{\parallel}$ vanishes. (b) The quantum
Fisher information per trial for $\hat{\mathcal{O}}_{\lambda}^{(0)}$
is consistently lower across all $\Theta$. (c) Scaled variance $\mathrm{Var}(\lambda)\times\mathcal{N}$
as a function of $\Theta$ for $j=5/2$. Both $\hat{\mathcal{O}}_{\lambda}^{P}$
and $\hat{\mathcal{O}}_{\lambda}^{(0)}$ exhibit increased variance
near $\Theta=0$, highlighting suboptimal performance in this regime,
while optimization using $\hat{\mathcal{O}}_{\lambda}^{P}$ yields
a lower variance at $\Theta_{\parallel}$.}\label{fig:5}
\end{figure*}

Thus, for $j=1/2$, the parallel
term $\mathcal{Q}^{\parallel}$, Eq. (\ref{eq:q=000020parallel2-1}), gives
\begin{equation}
\mathcal{Q}^{\parallel}=\frac{n^{2}}{16 d^{2}}\bigl|\mathcal{Z}{\scriptstyle (\lambda)}-e^{\imath\Theta}\mathcal{Z}_{0}\bigr|^{2},\label{eq:Q}
\end{equation}
where $\mathcal{Z}_{0}=d(d-1)/2$ and
\[
\mathcal{Z}{\scriptstyle (\lambda)}=\sum_{k=0}^{d-1}ke^{\imath nk\lambda}=\ensuremath{\frac{e^{\imath n\lambda}-de^{\imath nd\lambda}+(d-1)e^{\imath n(d+1)\lambda}}{\left(1-e^{\imath n\lambda}\right)^{2}}.}
\]
From Eq. (\ref{eq:Q}), the optimal postselection parameter $\Theta$ that satisfies the condition
$\mathcal{Q}^{\parallel}=0$, can be
expressed analytically as
\begin{equation}
\Theta_{\parallel}=-n\lambda+\mathrm{Im\,ln}\bigl[de^{\imath nd\lambda}-e^{\imath n\lambda}+(1-d)e^{\imath n(d+1)\lambda}\bigr],
\end{equation}
which defines the precise tuning condition for maximizing $\mathcal{T}$ for different values of $j$, as illustrated in Fig.~\ref{fig:5}(a). While $\mathcal{I}^{\parallel}$ is not strictly zero for $\Theta=\Theta_{\textrm{\ensuremath{\parallel}}}$ as in the case
$d=2$, Eq. (\ref{parrallel d=2}), its magnitude remains much smaller than the total term
$\mathcal{I}^{T}$, such that $\frac{\mathcal{I}^{\parallel}}{\mathcal{I}^{T}}\ll1.$This
indicates that the parallel component contributes negligibly and does
not affect the overall scaling of $\mathcal{I}^{T}$ for $\Theta=\Theta_{\textrm{\ensuremath{\parallel}}}$,
taking into account the final note in Section \ref{Complete Nullification}.

Figure  \ref{fig:4}(a) shows the postselected QFI as functions of $\Theta$ for $d=30$, illustrating
how $\hat{\mathcal{O}}_{\lambda}^{P}$ enables precise tuning of the
trade-off between orthogonal and parallel sensitivities. The results
in Fig.~\ref{fig:4}(b) reveal that the QFI associated with the operator $\hat{\mathcal{O}}_{\lambda}^{P}$
exhibits a pronounced upward scaling as the meter dimension $d$ increases.
Conversely, Fig.~\ref{fig:4}(c) shows that the QFI computed with the operator
$\hat{\mathcal{O}}_{\lambda}^{(0)}$ yields a QFI that is comparatively
insensitive to $d$. 

The performance comparison between $\hat{\mathcal{O}}_{\lambda}^{P}$ and
$\hat{\mathcal{O}}_{\lambda}^{(0)}$, Figure ~\ref{fig:5}(a) and ~\ref{fig:5}(b), demonstrates that the presence of Pancharatnam phase in the postselected system is crucial for unlocking the full advantage offered by increasing the meter quantum complexity in the compression of metrological information.

For $n\lambda=10^{-3}$ and $d=30$, under minimal loss conditions
with $\Theta=\Theta_{\textrm{\ensuremath{\parallel}}}$, the operator
$\hat{\mathcal{O}}_{\lambda}^{P}$ yields a postselection probability
of $25\times10^{-6}$. Achieving the same compression (i.e., equal
postselection probability) with the operator $\hat{\mathcal{O}}_{\lambda}^{(0)}$
requires $\Theta\simeq5.17n\lambda$. Consequently, for equivalent
compression rates, the metrological gain of $\hat{\mathcal{O}}_{\lambda}^{P}$
relative to $\hat{\mathcal{O}}_{\lambda}^{(0)}$ is 
\begin{equation}
g=10\log_{10}\frac{\mathcal{I}^{\perp}{\scriptstyle (\Theta_{\textrm{\ensuremath{\parallel}}})}\bigr|_{\hat{\mathcal{O}}_{\lambda}^{P}}}{\mathcal{I}^{\perp}{\scriptstyle (\Theta\simeq5.17n\lambda)}\bigr|_{\hat{\mathcal{O}}_{\lambda}^{(0)}}}=11.5\,\textrm{dB}
\end{equation}
for the same number of trials $\mathcal{N}$.

Figure ~\ref{fig:5}(c) shows that the variance $\textbf{Var}(\lambda)$ peaks near $\Theta=0$,
rendering this regime unsuitable for estimation.
In contrast, optimizing
at $\Theta_{\parallel}$ enables
saturation of the Cram\'{e}r-Rao bound (\ref{eq:=000020Cram=0000E9r-Rao=000020bound}),
thus achieving optimal measurement precision.

For $\Theta=\Theta_{\parallel}$, $j=1/2$ and in the large d limit, the QFI
per trial gives $\mathcal{T}{\scriptstyle (\lambda,\Theta_{\parallel})}=4\mathcal{Q}^{T}\simeq\frac{n^{2}d^{2}}{3}$.
 From Eq. (\ref{var with T}), the uncertainty becomes 
\begin{equation}
 \Delta \lambda  \geq\frac{\sqrt{3}}{nd}\frac{1}{\sqrt{\mathcal{N}}}.
\end{equation}
The protocol achieves Heisenberg-limited scaling with respect to both the number of probes and the dimensionality of the meter quantum state, demonstrating a sensitivity gain beyond classical and standard quantum strategies \cite{Giovannetti2005}.  Specifically, the metrological gain obtained by employing
a qudit GHZ state relative to a qubit GHZ state is quantified as $g=10\log_{10}\frac{\textrm{\textbf{Var}}(\lambda)_{\textrm{qudit\,GHZ}}}{\textrm{\textbf{Var}}(\lambda)_{\textrm{qubit\,GHZ}}}\approx10\log_{10}(d)$.
For $d=30$, this corresponds to a gain of approximately 15 dB.

The signal-to-noise ratio (SNR) quantifies the precision of the estimated
parameter $\lambda$ and satisfies the inequality 
\begin{align}
\textrm{SNR} & =\frac{\lambda}{\sqrt{\textrm{Var}(\lambda)}}\leq\lambda\sqrt{M\mathcal{I}^{\perp}{\scriptstyle (\Theta)}}=\lambda\sqrt{\mathcal{N}\mathcal{T}{\scriptstyle (\Theta)}}.
\end{align}
For the operator $\hat{\mathcal{O}}_{\lambda}^{P}$, and when $\Theta=\Theta_{\parallel}$,
the SNR gives 
\begin{equation}
\textrm{SNR}\leq\frac{nd\lambda}{\sqrt{3}}\sqrt{\mathcal{N}}=\frac{nd\lambda}{\sqrt{3}}\sqrt{\Gamma\tau},
\end{equation}
where $\Gamma$ is the generation rate of the quantum states, and
$\tau$ is the data acquisition time. This relationship captures the
fundamental dependence of measurement precision on both the quantum
resources and experimental parameters. Therefore, optimizing the parameters
$n$, $d$, and $\Gamma$ relative to $\lambda$ is vital to approach
or saturate the uncertainty limit $\Delta\lambda$.

These estimates delineate the fundamental minimal experimental thresholds required to attain the theoretical lower bound of precision. We emphasize that a rigorous analysis incorporating realistic noise models is essential in future work to accurately characterize the ultimate experimental limitations. Nonetheless, the geometric framework we establish remains fundamentally valid for optimizing protocol performance despite decoherence and operational imperfections that inevitably degrade achievable precision.

\subsection{Challenges and Key Considerations for Future Studies}

As shown in Figure \ref{fig:4}, the postselected QFI increases with increasing
d, with its peak shifting closer toward the Pancharatnam phase. This
increase, however, saturates around $d=30$. The question remains:
can we achieve further enhancement in the postselected QFI beyond
this point as $d$ continues to grow?

We delineate three key characteristic values of the postselection
parameter $\Theta$ 
\begin{align}
\Theta_{T}= & \underset{\Theta}{\textrm{arg}\,\textrm{max}\,}\mathcal{I}^{T}{\scriptstyle (\lambda,\Theta)}=\textrm{Im\,ln}\langle\hat{O}_{\lambda}^{P}\rangle,\\
\Theta_{\textrm{\ensuremath{\perp}}}= & \underset{\Theta}{\textrm{arg}\,\textrm{max}\,}\mathcal{I}^{\perp}{\scriptstyle (\lambda,\Theta)},\\
\Theta_{\parallel}= &  \underset{\Theta}{\textrm{arg}\,\textrm{min}\,}\mathcal{I}^{\parallel}{\scriptstyle (\lambda,\Theta)},
\end{align}
each marking distinct operational regimes, with $\Theta_{T}<\Theta_{\textrm{\ensuremath{\perp}}}<\Theta_{\parallel}$.
Achieving the most accessible enhancement in the quality of quantum
compression channels requires the suppression of parallel dynamics
precisely at the Pancharatnam phase. This condition leads to the equality
\begin{equation}
\Theta_{\parallel}=\Theta_{\textrm{\ensuremath{\perp}}}=\textrm{Im\,ln}\langle\hat{O}_{\lambda}^{P}\rangle.\label{eq:equality}
\end{equation}
Future research should focus on manipulating the Pancharatnam phase
and exploring meter operators capable of achieving this critical equality.
Detailed investigation into the structural and operational properties
of such meter operators is essential to optimize many features of
postselected quantum metrology protocols (Table \ref{tab:1} summarizes the broad spectrum of quantum states considered in this study. Further investigations may directly incorporate other meter states, such as squeezed  states \citep{Lvovsky} and entangled coherent  states \citep{Sanders_2012}).
\begin{figure}
\begin{centering}
\includegraphics[width=0.85\columnwidth]{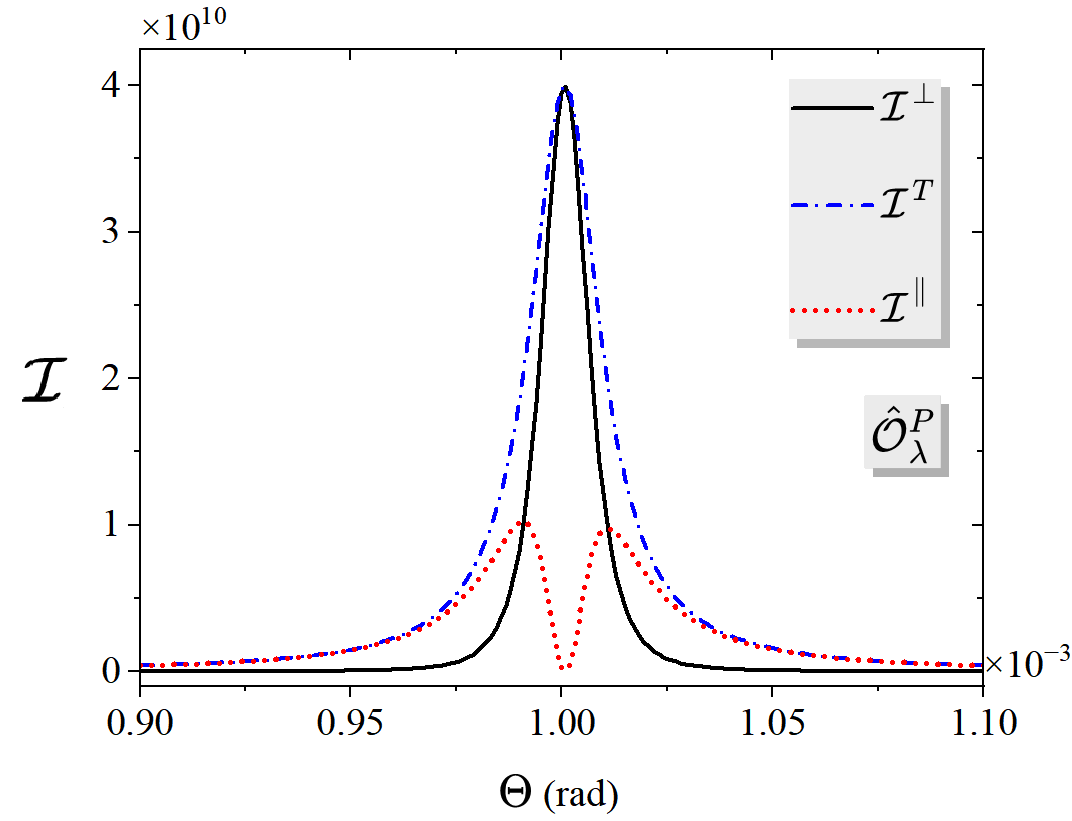}
\par\end{centering}
\caption{QFI $\mathcal{I}^{\perp}$, $\mathcal{I}^{T}$,
and $\mathcal{I}^{\parallel}$ are plotted as functions
of the postselection parameter $\Theta$ for $\lambda=10^{-3}$,
$d=10^{4}$,  $\varepsilon=10^{-4}$, and arbitrary $j$. The results demonstrate a pronounced
central peak in the total and orthogonal components, signifying optimal
quantum sensitivity at the corresponding postselection parameter $\Theta$.
The parallel component, by contrast, exhibits a split-peak structure
and vanishes at the central phase point, illustrating the complete
suppression of unobservable parallel evolution at the Pancharatnam
phase given in Eq. (\ref{eq:Panchartanam=000020phase=000020f}). }\label{fig:6}
\end{figure}

To illustrate this point, consider the meter operator 
\begin{equation}
\ensuremath{\hat{\mathcal{O}}_{\lambda}^{P}=\sum_{k=0}^{d-1}e^{\imath k^{\varepsilon}\lambda}|k\rangle\langle k|,}\label{eq:OPERATOR}
\end{equation}
 with the fractional exponent approaching zero, $\varepsilon\rightarrow0^{+}$,
representing a slowly varying phase. This form leads to the expectation
value
\begin{equation}
\langle\hat{O}_{\lambda}^{P}\rangle=\frac{1}{d}\sum_{k=0}^{d-1}e^{\imath\lambda k^{\varepsilon}}.\label{eq:=000020enhanced=000020operator}
\end{equation}
This sum is approximated as $\sum_{k=0}^{d-1}e^{\imath\lambda k^{\varepsilon}}\approx1+e^{\imath\lambda}\sum_{k=1}^{d-1}k^{\imath\lambda\varepsilon},$
where $k^{\imath\lambda\varepsilon}=e^{\imath\lambda\varepsilon\textrm{ln}(k)}.$
Using the integral approximation for large $d$ we get \footnote{In this regime of small-$\varepsilon$ and large-$d$ limits, $\mathcal{T}{\lambda,\scriptstyle (\Theta)}$
asymptotically approaches unity, presenting a nuanced challenge when
employing such mathematical operators in Eq. (\ref{eq:OPERATOR}).} 
\begin{equation}
\sum_{k=1}^{d-1}k^{\imath\lambda\varepsilon}\approx\int_{1}^{d}x^{\imath\lambda\varepsilon}dx=\frac{d^{1+\imath\lambda\varepsilon}-1}{1+\imath\lambda\varepsilon},
\end{equation}
 thus $\sum_{k=0}^{d-1}e^{\imath\lambda k^{\varepsilon}}\approx1+e^{\imath\lambda}\frac{d^{1+\imath\lambda\varepsilon}-1}{1+\imath\lambda\varepsilon},$
and the expectation value can be rewritten as
\begin{equation}
\langle\hat{O}_{\lambda}^{P}\rangle\approx\frac{1}{d}\bigl[1+e^{\imath\lambda}\frac{d^{1+\imath\lambda\varepsilon}-1}{1+\imath\lambda\varepsilon}\bigr]=\frac{1}{d}+e^{\imath\lambda}\frac{d^{\imath\lambda\varepsilon}-d^{-1}}{1+\imath\lambda\varepsilon}.
\end{equation}
where $d^{\imath\lambda\varepsilon}=e^{\imath\lambda\varepsilon\textrm{ln}(d)}\approx1+\imath\lambda\varepsilon\textrm{ln(}d)+\ldots$,
using the condition $|\lambda\varepsilon\textrm{ln}(d)|\ll1$, and $\frac{1}{1+\imath\lambda\varepsilon}\approx1-\imath\lambda\varepsilon-\lambda^{2}\varepsilon^{2}+\ldots$.
This gives
\begin{equation}
\langle\hat{O}_{\lambda}^{P}\rangle\approx e^{\imath\lambda}\bigl\{1+\imath\lambda\varepsilon[\textrm{ln(}d)-1]\bigr\}+\frac{1-e^{\imath\lambda}}{d}+\ldots.
\end{equation}
Noting that $\textrm{ln}(1+\delta)\approx\delta$ for small $\delta$,
where $\delta=\imath\lambda\varepsilon[\textrm{ln(}d)-1]+\frac{e^{-\imath\lambda}-1}{d}+\ldots.$
The Pancharatnam phase gives
\begin{align}
\textrm{Im\,ln}\langle\hat{O}_{\lambda}^{P}\rangle & =\lambda+\lambda\varepsilon\bigl[\textrm{ln(}d)-1\bigr]-\frac{\sin\lambda}{d}+\ldots.\nonumber \\
 & \approx\lambda\bigl[1-\varepsilon+\varepsilon\textrm{ln}(d)\bigr]\approx\lambda.\label{eq:Panchartanam=000020phase=000020f}
\end{align}
As illustrated in Fig. \ref{fig:6}, in both the small-$\varepsilon$ and large-$d$
limits, the equality (\ref{eq:equality}), holds. At this critical
phase, the unobservable parallel evolution is completely eliminated,
i.e., $\mathcal{I}^{\parallel}\approx 0$,
while the compression quality achieves its maximum by fully concentrating
all accessible information contained in $\mathcal{I}^{T}$
into the experimentally measurable orthogonal component $\mathcal{I}^{\perp}$.
This demonstrates the Pancharatnam phase acts as a natural, robust
benchmark for optimizing compression quality of quantum channels.

\section{Conclusion}

We identify the noncyclic Pancharatnam phase, arising from coherent system-meter coupling, as the central geometric criterion for optimal quantum compression across a broad class of postselected metrology protocols.

The realization of a truly lossless compression channel is shown to be contingent upon two principles: the system's evolution must be constrained to a Pancharatnam phase (nonvanishing connection), and the transition from the pre- to postselected state must constitute a single quantum.

This study reveals that slight deviations from the Pancharatnam phase cause pronounced reductions in both compression efficiency and metrological precision, highlighting its practical role in experimental realizations. 

Significantly, our framework further reveals that exploiting the extended complexity of quantum meters--specifically through high-dimensional qudit systems--offers a pathway to unlock substantial additional gains in quantum compression and measurement sensitivity.

The Pancharatnam phase thereby constitutes a foundational benchmark, advancing the theoretical foundations of quantum parameter estimation in postselected metrology.

\begin{acknowledgments}
The authors thank Prof. Leonid Il'ichov for valuable discussions.
This work was carried out at the Institute of Automation and Electrometry SB RAS  under the framework of the State Assignment (project 124041700105-5)
.
\end{acknowledgments}

\appendix

\section{Quantum Parallel Transport Condition\protect\label{sec:Quantum-parallel-transport}}

A canonical example of parallel transport in classical physics is
the transport of a tangent vector along a closed loop on the surface
of a sphere. Consider a vector initially defined at point $A$ and
transported along the geodesic triangle $A\rightarrow B\rightarrow C\rightarrow A$.
Throughout this process, the vector remains confined to the tangent
plane of the sphere, and no local rotation (phase accumulation) occurs
along the geodesic segments $AB$, $BC$, and $CA$. Upon returning
to the initial point $A$, the vector generally undergoes a net rotation
relative to its original orientation, a manifestation of the sphere\textquoteright s
intrinsic curvature encoded by the holonomy associated with the loop.

\begin{figure}[t]
\begin{centering}
\includegraphics[width=0.7\columnwidth]{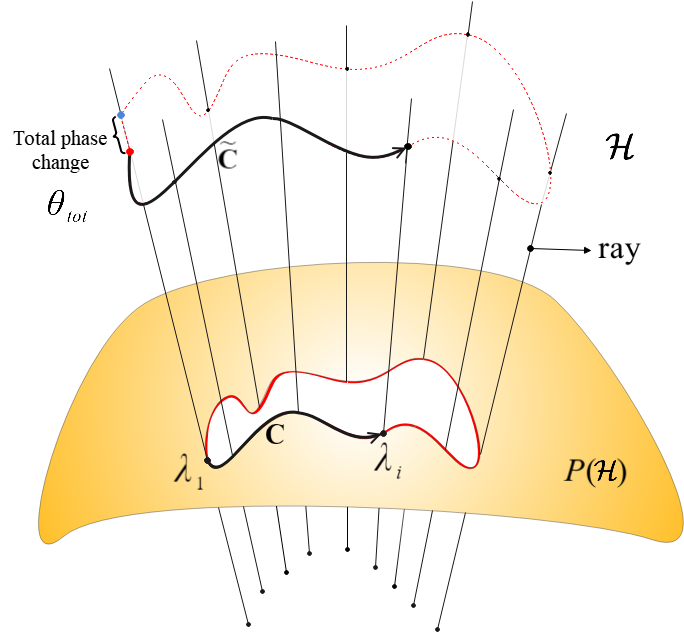}
\par\end{centering}
\caption{The set of all equivalent state vectors (a ray) represents a point
in the projective Hilbert  space $P\scriptstyle(\mathcal{H})$. }\label{fig:8}
\end{figure}

By analogy, the concept of parallel transport extends naturally to
the evolution of quantum states in a Hilbert space $\mathcal{H}$.
Quantum states differing only by a global phase factor form equivalence
classes known as rays, which correspond to one-dimensional subspaces
spanned by normalized vectors. The collection of all such rays constitutes
the projective Hilbert space $P\scriptstyle(\mathcal{H})$, a manifold
that effectively removes the physically irrelevant global phase from
the state description; see Fig.~\ref{fig:8}. This space can be identified with the set of
pure-state density matrices modulo global phase, serving as the natural
arena for geometric phases and holonomies in quantum mechanics.

The quantum parallel transport condition can be rigorously formulated
to ensure the absence of any geometric phase accumulation as the parameter
$\lambda$ varies infinitesimally \cite{Anandan1992}. Concretely, this condition requires
that no relative phase is introduced between the states $\left|\Psi{\scriptstyle (\lambda)}\right\rangle $
and $\left|\Psi{\scriptstyle(\lambda+d\lambda)}\right\rangle $, which is mathematically
equivalent to demanding that their overlap $\left\langle \Psi{\scriptstyle (\lambda)}\left|\Psi{\scriptstyle(\lambda+d\lambda)}\right.\right\rangle $
remains strictly real and positive. The argument of this overlap,
defined as $\arg\left\langle \Psi{\scriptstyle (\lambda)}\left|\Psi{\scriptstyle(\lambda+d\lambda)}\right.\right\rangle =\textrm{Im\,ln}\left\langle \Psi{\scriptstyle (\lambda)}\left|\Psi{\scriptstyle(\lambda+d\lambda)}\right.\right\rangle $
vanishes identically. Accordingly
\begin{align*}
\textrm{Im\,Ln}\left\langle \Psi{\scriptstyle (\lambda)}\left|\Psi{\scriptstyle(\lambda+d\lambda)}\right.\right\rangle =\, & \textrm{Im\,Ln}\bigl(\langle\Psi{\scriptstyle (\lambda)}|\Psi{\scriptstyle (\lambda)}\rangle\\
 & +\langle\Psi{\scriptstyle (\lambda)}|\partial_{\lambda}\Psi{\scriptstyle (\lambda)}\rangle d\lambda+O(d\lambda^{2})\bigr)\\
\simeq\, & \textrm{Im\,Ln}\bigl(1+\langle\Psi{\scriptstyle (\lambda)}|\partial_{\lambda}\Psi{\scriptstyle (\lambda)}\rangle d\lambda\bigr)\\
\simeq\, & \langle\Psi{\scriptstyle (\lambda)}|\partial_{\lambda}\Psi{\scriptstyle (\lambda)}\rangle d\lambda=0,
\end{align*}
and the quantum parallel transport condition can be expressed by $\langle\Psi{\scriptstyle (\lambda)}|\partial_{\lambda}\Psi{\scriptstyle (\lambda)}\rangle=0$,
which imposes that the tangent vector $|\partial_{\lambda}\Psi{\scriptstyle (\lambda)}\rangle$
remains orthogonal to the state vector $|\Psi{\scriptstyle (\lambda)}\rangle$ in
Hilbert space, i.e., $|\partial_{\lambda}\Psi{\scriptstyle (\lambda)}\rangle\perp|\Psi{\scriptstyle (\lambda)}\rangle.$
This orthogonality condition defines a horizontal lift of the parameterized
curve from the projective Hilbert space to the full Hilbert space,
eliminating gauge-dependent dynamical phase contributions. 
\begin{table*}[t]
\begin{centering}
\begin{tabular}{c|c|llc|c}
\hline 
\multirow{1}{*}{\textbf{Number of Parties (n)}} & \multirow{1}{*}{\textbf{Dimension (d)}} &  &  & \textbf{Meter state $|\mathrm{\mathcal{M}}_{d}^{(n)}\rangle$} & \textbf{State Classification}\tabularnewline
\hline 
\multicolumn{1}{c}{} & \multicolumn{1}{c}{} &  &  & \multicolumn{1}{c}{} & \tabularnewline
\hline 
\multirow{4}{*}{1} & 1 &  &  & $|b_{0}\rangle$ & basis\tabularnewline
\cline{2-5}
 & 2 &  &  & $\frac{1}{\sqrt{2}}(|b_{0}\rangle+|b_{1}\rangle)$ & qubit\tabularnewline
\cline{2-5}
 & 3 &  &  & $\frac{1}{\sqrt{3}}(|b_{0}\rangle+|b_{1}\rangle+|b_{2}\rangle)$ & qutrit\tabularnewline
\cline{2-5}
 & d &  &  & $\frac{1}{\sqrt{d}}(|b_{0}\rangle+|b_{1}\rangle+\ldots+|b_{d-1}\rangle)$ & qudit\tabularnewline
\hline 
\multicolumn{6}{c}{}\tabularnewline
\hline 
\multirow{4}{*}{2} & 1 &  &  & $|b_{0}\rangle|b_{0}\rangle,$ & bipartite product (twin)\tabularnewline
\cline{2-5}
 & 2 &  &  & $\frac{1}{\sqrt{2}}(|b_{0}\rangle|b_{0}\rangle+|b_{1}\rangle|b_{1}\rangle)$ & maximally entangled two-qubit\tabularnewline
\cline{2-5}
 & 3 &  &  & $\frac{1}{\sqrt{3}}(|b_{0}\rangle|b_{0}\rangle+|b_{1}\rangle|b_{1}\rangle+|b_{2}\rangle|b_{2}\rangle)$ & maximally entangled two-qutrit\tabularnewline
\cline{2-5}
 & d &  &  & $\frac{1}{\sqrt{d}}(|b_{0}\rangle|b_{0}\rangle+|b_{1}\rangle|b_{1}\rangle+\ldots+|b_{d-1}\rangle|b_{d-1}\rangle)$ & maximally entangled two-qudit\tabularnewline
\hline 
\multicolumn{6}{c}{}\tabularnewline
\hline 
\multirow{4}{*}{3} & 1 &  &  & $|b_{0}\rangle|b_{0}\rangle|b_{0}\rangle$ & tripartite product\tabularnewline
\cline{2-5}
 & 2 &  &  & $\frac{1}{\sqrt{2}}(|b_{0}\rangle|b_{0}\rangle|b_{0}\rangle+|b_{1}\rangle|b_{1}\rangle|b_{1}\rangle)$ & GHZ\tabularnewline
\cline{2-5}
 & 3 &  &  & $\frac{1}{\sqrt{3}}(|b_{0}\rangle|b_{0}\rangle|b_{0}\rangle+|b_{1}\rangle|b_{1}\rangle|b_{1}\rangle+|b_{2}\rangle|b_{2}\rangle|b_{2}\rangle)$ & qutrit (generalized) GHZ\tabularnewline
\cline{2-5}
 & d &  &  & $\frac{1}{\sqrt{d}}(|b_{0}\rangle|b_{0}\rangle|b_{0}\rangle+|b_{1}\rangle|b_{1}\rangle|b_{1}\rangle\ldots+|b_{d-1}\rangle|b_{d-1}\rangle|b_{d-1}\rangle)$ & qudit (generalized) GHZ\tabularnewline
\hline 
\multicolumn{1}{c}{} & \multicolumn{1}{c}{} &  &  & \multicolumn{1}{c}{} & \tabularnewline
\hline 
n & d &  &  & $\frac{1}{\sqrt{d}}(|b_{0}\rangle^{\otimes n}+|b_{1}\rangle^{\otimes n}+\ldots+|b_{d-1}\rangle^{\otimes n})$ & n-partite qudit (generalized) GHZ\tabularnewline
\hline 
\end{tabular}
\par\end{centering}
\caption{\protect\label{tab:1}Meter state $|\mathrm{\mathcal{M}}_{d}^{(n)}\rangle$
classification for different values of d and n.}
\end{table*}

\section{Geometrical Description of  QFI\protect\label{subsec:Geometrical-description-of}}

When a quantum state $|\Psi{\scriptstyle (\lambda)}\rangle$ exhibits parametric
dependence on $\lambda$, its derivative $|\partial_{\lambda}\Psi{\scriptstyle (\lambda)}\rangle$
describes the state infinitesimal variations. These variations encompass
two distinct contributions: $(i)$ a physical component reflecting
the state\textquoteright s evolution to a different ray within the
projective Hilbert space, and $(ii)$ a gauge-dependent component
corresponding to an infinitesimal phase rotation, which is parallel
to $|\Psi{\scriptstyle (\lambda)}\rangle$. Thus, the derivative with respect to
the parameter $\lambda$ decomposes as:
\begin{equation}
|\partial_{\lambda}\Psi{\scriptstyle (\lambda)}\rangle=|\partial_{\lambda}\Psi{\scriptstyle (\lambda)}\rangle^{\parallel}+|\partial_{\lambda}\Psi{\scriptstyle (\lambda)}\rangle^{\perp}.\label{eq:decomposition}
\end{equation}
The component of $|\partial_{\lambda}\Psi{\scriptstyle (\lambda)}\rangle$ parallel
to $\left|\Psi{\scriptstyle (\lambda)}\right\rangle $ is given by \cite{PhysRevLett.72.3439}
\[
|\partial_{\lambda}\Psi{\scriptstyle (\lambda)}\rangle^{\parallel}=\mathbb{P}_{\psi}\left|\!\right.\partial_{\lambda}\Psi{\scriptstyle (\lambda)}\left.\!\right\rangle =\left|\Psi{\scriptstyle (\lambda)}\right\rangle \left\langle \Psi{\scriptstyle (\lambda)}\right|\partial_{\lambda}\Psi{\scriptstyle (\lambda)}\left.\!\right\rangle ,
\]
where $\mathbb{P}_{\psi}=\left|\Psi{\scriptstyle (\lambda)}\right\rangle \left\langle \Psi{\scriptstyle (\lambda)}\right|$
is the projector onto the state $\left|\Psi{\scriptstyle (\lambda)}\right\rangle $.
The overlap $\left\langle \Psi{\scriptstyle (\lambda)}\right|\partial_{\lambda}\Psi{\scriptstyle (\lambda)}\left.\!\right\rangle $
causes a global phase change that does not affect measurement outcomes.
The orthogonal component is derived simply by subtraction of the parallel
component from $|\partial_{\lambda}\Psi{\scriptstyle (\lambda)}\rangle$ and can
be written as \cite{PhysRevLett.72.3439}
\begin{align}
|\partial_{\lambda}\Psi{\scriptstyle (\lambda)}\rangle^{\perp}=\,\, & |\partial_{\lambda}\Psi{\scriptstyle (\lambda)}\rangle-\left|\negmedspace\right.\partial_{\lambda}\Psi{\scriptstyle (\lambda)}\left.\negmedspace\right\rangle ^{\parallel}\nonumber \\
=\,\, & (\mathbb{I}-|\Psi{\scriptstyle (\lambda)}\rangle\langle\Psi{\scriptstyle (\lambda)}|)|\partial_{\lambda}\Psi{\scriptstyle (\lambda)}\rangle,
\end{align}
which leads to observable differences in the quantum state. The decomposition
(\ref{eq:decomposition}) of the state derivative becomes 
\begin{equation}
|\partial_{\lambda}\Psi{\scriptstyle (\lambda)}\rangle=\negthickspace\underbrace{\langle\Psi{\scriptstyle (\lambda)}|\partial_{\lambda}\Psi{\scriptstyle (\lambda)}\rangle|\Psi{\scriptstyle (\lambda)}\rangle}_{\text{Parallel (global phase)}}\negthickspace+\negthinspace\underbrace{(\mathbb{I}-|\Psi{\scriptstyle (\lambda)}\rangle\langle\Psi{\scriptstyle (\lambda)}|)|\partial_{\lambda}\Psi{\scriptstyle (\lambda)}\rangle}_{\text{Orthogonal (physical)}},
\end{equation}
which clearly isolates the physically relevant sensitivity of the state
to $\lambda$. The QFI can then be written as \cite{Yang2024,ArvidssonShukur2020}
\begin{align}
\mathcal{I}^{\perp}= & 4\bigr\Vert|\partial_{\lambda}\Psi{\scriptstyle (\lambda)}\rangle^{\perp}\bigr\Vert^{2}\nonumber \\
= & 4\bigl(\langle\partial_{\lambda}\Psi{\scriptstyle (\lambda)}|\partial_{\lambda}\Psi{\scriptstyle (\lambda)}\rangle-|\langle\Psi{\scriptstyle (\lambda)}|\partial_{\lambda}\Psi{\scriptstyle (\lambda)}\rangle|^{2}\bigr)\nonumber\\
= & 4(\bigr\Vert|\partial_{\lambda}\Psi{\scriptstyle (\lambda)}\rangle\bigr\Vert^{2}-\bigr\Vert|\partial_{\lambda}\Psi{\scriptstyle (\lambda)}\rangle^{\parallel}\bigr\Vert^{2}),
\end{align}
which rigorously quantifies the parametric sensitivity of a quantum
state by evaluating the squared norm of the component of the state\textquoteright s
derivative that is orthogonal to the state vector itself in Hilbert
space. This orthogonal component encapsulates the physically meaningful
variation of the state within the projective Hilbert space, effectively
filtering out gauge-dependent phase fluctuations parallel to the state
vector.

\section{Gauge Invariance of QFI\label{Gauge invariance}}

By performing a phase redefinition (gauge transformation) for the
quantum state
\[
\ensuremath{|\Psi{\scriptstyle (\lambda)}\rangle\to e^{\imath\varphi{\scriptstyle (\lambda)}}|\Psi{\scriptstyle (\lambda)}\rangle,}
\]
where $\varphi{\scriptstyle (\lambda)}$ is an arbitrary real function, the total derivative
reads
\[
|\partial_{\lambda}\Psi{\scriptstyle (\lambda)}\rangle\rightarrow e^{\imath\varphi{\scriptstyle (\lambda)}}(\imath|\Psi{\scriptstyle (\lambda)}\rangle\partial_{\lambda}\varphi{\scriptstyle (\lambda)}+|\partial_{\lambda}\Psi{\scriptstyle (\lambda)}\rangle),
\]
and its norm gives
\begin{align}
\bigr\Vert|\partial_{\lambda}\Psi{\scriptstyle (\lambda)}\rangle\bigr\Vert^{2}\to & \,\,\langle\partial_{\lambda}\Psi{\scriptstyle (\lambda)}|\partial_{\lambda}\Psi{\scriptstyle (\lambda)}\rangle+(\partial_{\lambda}\varphi{\scriptstyle (\lambda)})^{2}\nonumber \\
 & +2\imath\partial_{\lambda}\varphi{\scriptstyle (\lambda)}\langle\Psi{\scriptstyle (\lambda)}|\partial_{\lambda}\Psi{\scriptstyle (\lambda)}\rangle,
\end{align}
where $\langle\Psi{\scriptstyle (\lambda)}|\Psi{\scriptstyle (\lambda)}\rangle=1.$ The parallel
term transforms as
\begin{align}
\bigr\Vert|\partial_{\lambda}\Psi{\scriptstyle (\lambda)}\rangle^{\parallel}\bigr\Vert^{2}\to & \,\,|\langle\Psi{\scriptstyle (\lambda)}|\partial_{\lambda}\Psi{\scriptstyle (\lambda)}\rangle+\imath\partial_{\lambda}\varphi{\scriptstyle (\lambda)}|^{2}\nonumber \\
= & \,\,|\langle\Psi{\scriptstyle (\lambda)}|\partial_{\lambda}\Psi{\scriptstyle (\lambda)}\rangle|^{2}+(\partial_{\lambda}\varphi{\scriptstyle (\lambda)})^{2}\nonumber \\
 & +2\imath\partial_{\lambda}\varphi{\scriptstyle (\lambda)}\langle\Psi{\scriptstyle (\lambda)}|\partial_{\lambda}\Psi{\scriptstyle (\lambda)}\rangle.
\end{align}
Substituting the last two expressions into the QFI formula (\ref{eq:standard=00003D000020qfi}),
we find that the quantities involving $\varphi{\scriptstyle (\lambda)}$ cancel exactly
\begin{align}
\mathcal{I}^{\perp}= & \,\,4\langle\partial_{\lambda}\Psi{\scriptstyle (\lambda)}|\partial_{\lambda}\Psi{\scriptstyle (\lambda)}\rangle-4|\langle\Psi{\scriptstyle (\lambda)}|\partial_{\lambda}\Psi{\scriptstyle (\lambda)}\rangle|^{2}.
\end{align}
The QFI formula is gauge-invariant.

\section{Derivation of QFI Associated with $\left|\Psi{\scriptstyle (\lambda)}\right\rangle $\protect\label{sec:Derivation-of-the}}

To derive the QFI formula, we start from the compressed state \cite{Yang2024} 
\begin{equation}
\left|\Psi{\scriptstyle (\lambda)}\right\rangle =\mathcal{P}^{-1/2}{\scriptstyle (\lambda)}|\mathcal{S}_{f}\rangle\otimes\hat{\mathds{K}}{\scriptstyle (\lambda)}|\mathcal{M}_{i}\rangle=|\mathcal{S}_{f}\rangle\otimes\left|\Phi{\scriptstyle (\lambda)}\right\rangle ,\label{eq:meter=000020state=0000201}
\end{equation}
and find the derivative
\begin{equation}
|\partial_{\lambda}\Phi{\scriptstyle (\lambda)}\rangle=\mathcal{P}^{-1/2}{\scriptstyle (\lambda)}\bigl[\partial_{\lambda}\hat{\mathds{K}}{\scriptstyle (\lambda)}-\tfrac{1}{2}\hat{\mathds{K}}{\scriptstyle (\lambda)}\partial_{\lambda}\textrm{Log}\mathcal{P}{\scriptstyle (\lambda)}\bigr]\left|\!\right.\mathcal{M}_{i}\left.\!\right\rangle .\label{eq:meter=000020state}
\end{equation}
Thus, the magnitude of the total derivative is given by
\begin{align}
\bigr\Vert|\partial_{\lambda}\Psi{\scriptstyle (\lambda)}\rangle\bigr\Vert^{2}= & \,\,\mathcal{P}^{-1}{\scriptstyle (\lambda)}\bigl[\left\langle \negmedspace\right.\partial_{\lambda}\hat{\mathds{K}}^{\dagger}{\scriptstyle (\lambda)}\partial_{\lambda}\hat{\mathds{K}}{\scriptstyle (\lambda)}\left.\negmedspace\right\rangle \nonumber \\
 & -\tfrac{\mathcal{P}^{-1}{\scriptstyle (\lambda)}}{2}\left\langle \negmedspace\right.\partial_{\lambda}(\hat{\mathds{K}}^{\dagger}{\scriptstyle (\lambda)}\hat{\mathds{K}}{\scriptstyle (\lambda)})\left.\negmedspace\right\rangle \partial_{\lambda}\textrm{Log}\mathcal{P}{\scriptstyle (\lambda)}\bigr]\nonumber \\
 & +\tfrac{1}{4}(\partial_{\lambda}\textrm{Log}\mathcal{P}{\scriptstyle (\lambda)})^{2}.
\end{align}
Using expressions (\ref{eq:meter=000020state=0000201}) and (\ref{eq:meter=000020state})
one can find
\begin{align}
\left\langle \!\right.\Phi{\scriptstyle (\lambda)}|\partial_{\lambda}\Phi{\scriptstyle (\lambda)}\rangle= & \mathcal{P}^{-1}{\scriptstyle (\lambda)}\left\langle \negmedspace\right.\hat{\mathds{K}}^{\dagger}{\scriptstyle (\lambda)}\partial_{\lambda}\hat{\mathds{K}}{\scriptstyle (\lambda)}\left.\negmedspace\right\rangle \nonumber \\
 & -\tfrac{1}{2}\partial_{\lambda}\textrm{Log}\mathcal{P}{\scriptstyle (\lambda)},\\
\left\langle \!\right.\partial_{\lambda}\Phi{\scriptstyle (\lambda)}|\Phi{\scriptstyle (\lambda)}\rangle= & \mathcal{P}^{-1}{\scriptstyle (\lambda)}\left\langle \negmedspace\right.\partial_{\lambda}\hat{\mathds{K}}^{\dagger}{\scriptstyle (\lambda)}\hat{\mathds{K}}{\scriptstyle (\lambda)}\left.\negmedspace\right\rangle \nonumber \\
 & -\tfrac{1}{2}\partial_{\lambda}\textrm{Log}\mathcal{P}{\scriptstyle (\lambda)},
\end{align}
and the parallel term in the QFI gives
\begin{align}
\bigr\Vert|\partial_{\lambda}\Psi{\scriptstyle (\lambda)}\rangle^{\parallel}\bigr\Vert^{2}= & \mathcal{P}^{-2}{\scriptstyle (\lambda)}\left|\negthickspace\right.\left\langle \!\right.\hat{\mathds{K}}^{\dagger}{\scriptstyle (\lambda)}\,\partial_{\lambda}\hat{\mathds{K}}{\scriptstyle (\lambda)}\left.\!\right\rangle \left.\negthickspace\right|^{2}\nonumber \\
 & -\tfrac{\mathcal{P}^{-1}{\scriptstyle (\lambda)}}{2}\left\langle \negmedspace\right.\partial_{\lambda}(\hat{\mathds{K}}^{\dagger}{\scriptstyle (\lambda)}\hat{\mathds{K}}{\scriptstyle (\lambda)})\left.\negmedspace\right\rangle \partial_{\lambda}\textrm{Log}\mathcal{P}{\scriptstyle (\lambda)}\nonumber \\
 & +\tfrac{1}{4}(\partial_{\lambda}\textrm{Log}\mathcal{P}{\scriptstyle (\lambda)})^{2}.
\end{align}
The final form of the QFI becomes
\begin{align}
\mathcal{I}^{\perp}= & 4\bigl[\bigr\Vert|\partial_{\lambda}\Psi{\scriptstyle (\lambda)}\rangle\bigr\Vert^{2}-\bigr\Vert|\partial_{\lambda}\Psi{\scriptstyle (\lambda)}\rangle^{\parallel}\bigr\Vert^{2}\bigr]\nonumber \\
= & 4\mathcal{P}^{-2}{\scriptstyle (\lambda)}\bigl[\left\langle \negmedspace\right.\partial_{\lambda}\hat{\mathds{K}}^{\dagger}{\scriptstyle (\lambda)}\partial_{\lambda}\hat{\mathds{K}}{\scriptstyle (\lambda)}\left.\negmedspace\right\rangle \mathcal{P}{\scriptstyle (\lambda)} -\left|\negthickspace\right.\left\langle \!\right.\hat{\mathds{K}}^{\dagger}{\scriptstyle (\lambda)}\,\partial_{\lambda}\hat{\mathds{K}}{\scriptstyle (\lambda)}\left.\!\right\rangle \left.\negthickspace\right|^{2}\bigr],
\end{align}
which is the QFI written as a function of the channel operator $\hat{\mathds{K}}{\scriptstyle (\lambda)}$.

We note that the averages in all previous calculations are
evaluated with respect to the meter state, which is defined as the
qudit state $|\mathcal{M}_{i}\rangle=|\mathrm{\mathcal{M}}_{d}^{(n)}\rangle=\frac{1}{\sqrt{d}}\sum_{k=0}^{d-1}|b_{k}\rangle^{\otimes n}$, as
introduced in Section \ref{Qudit Meter State}. The corresponding states for different values
of n and d are listed in Table  \ref{tab:1}.

\section{Noncyclic Pancharatnam Phase\protect\label{sec:Pancharatnam-phase}}

To elucidate the nature of the Pancharatnam phase in Eq. (\ref{eq:pancharatnam=000020shift}),
we exploit its characteristic non-transitivity \citep{Berry1987},
which directly reflects the geometric origin of the phase difference.
Specifically, we consider the evolution generated by the operator
$\hat{\mathcal{O}}_{\lambda}$ as a sequence of $i$ discrete transitions
between neighboring states, defined by parameters $\lambda$ with
$\lambda_{1}=d\lambda,\lambda_{2}=\lambda_{1}+d\lambda,\ldots,$$\lambda_{i}=\lambda_{i-1}+d\lambda$.
The phase difference between the initial and final states is then
constructed via successive projections: the initial state $|\phi_{1}\rangle$
is projected onto the second state $|\mathrm{\mathrm{\phi_{2}}}\rangle\langle\phi_{2}|\phi_{1}\rangle$,
the second onto the third, and so forth, until the final projection
onto the original state $|\phi_{1}\rangle\langle\phi_{1}|\phi_{i}\rangle$.
Consequently, the final state is expressed as \cite{Mukunda1993}
\begin{equation}
|\widetilde{\phi}_{1}\rangle=|\phi_{1}\rangle\langle\phi_{1}|\phi_{i}\rangle\langle\phi_{i}|\ldots\langle\phi_{4}|\phi_{3}\rangle\langle\phi_{3}|\phi_{2}\rangle\langle\phi_{2}|\phi_{1}\rangle,\label{eq:9}
\end{equation}
which differs from the initial state $|\phi_{1}\rangle$, by the phase
factor 
\begin{align}
\theta_{geo}= & \textrm{Im\,ln}\left[\langle\phi_{1}|\phi_{i}\rangle\langle\phi_{i}|\ldots\langle\phi_{4}|\phi_{3}\rangle\langle\phi_{3}|\phi_{2}\rangle\langle\phi_{2}|\phi_{1}\rangle\right]\nonumber \\
= & \textrm{Im\,ln}\left[\textrm{Tr}\bigl(|\phi_{i}\rangle\langle\phi_{i}|\ldots\langle\phi_{3}|\phi_{2}\rangle\langle\phi_{2}|\phi_{1}\rangle\langle\phi_{1}|\bigr)\right]\nonumber \\
= & \textrm{Im\,ln}\bigl[\textrm{Tr}\prod_{k=1}^{i}\varrho_{k}\bigr]
\end{align}
corresponds to the geometric phase, which is a gauge-invariant quantity
for cyclic evolution, where the initial and final states coincide.
In contrast, for noncyclic evolution \citep{Mukunda1993}, the final
state $|\phi_{i}\rangle$, does not project back onto the initial
state $|\phi_{1}\rangle$, and the geometric phase can be generalized
accordingly as
\begin{equation}
\begin{aligned}\theta_{geo} & =\textrm{Im\,ln}\left[\langle\phi_{1}|\phi_{i}\rangle\langle\phi_{i}|\ldots\langle\phi_{4}|\phi_{3}\rangle\langle\phi_{3}|\phi_{2}\rangle\langle\phi_{2}|\phi_{1}\rangle\right]\\
 & =\textrm{Im\,ln}\langle\phi_{1}|\phi_{i}\rangle-\textrm{\textrm{Im\,ln}}\prod_{k=1}^{i-1}\langle\phi_{k}|\phi_{k+1}\rangle\\
 & \approx\textrm{Im\,ln}\langle\phi_{1}|\phi_{i}\rangle-\textrm{Im}\sum_{k=1}^{i-1}\textrm{ln}\left[1+\bigl\langle\phi_{k}|\partial_{\lambda}\phi_{k}\bigr\rangle d\lambda+\ldots\right]\\
 & \approx\textrm{Im\,ln}\langle\phi_{1}|\hat{\mathcal{O}}_{\lambda}|\phi_{1}\rangle-\textrm{Im}\sum_{k=1}^{i-1}\bigl\langle\phi_{k}|\partial_{\lambda}\phi_{k}\bigr\rangle d\lambda,
\end{aligned}
\end{equation}
where $\textrm{Im\,ln}\langle\phi_{1}|\phi_{i}\rangle=\textrm{Im\,ln}\langle\hat{\mathcal{O}}_{\lambda}\rangle$
is the total phase difference between initial and final states. The
second term involving $\textrm{Im}\sum_{k=1}^{i-1}\bigl\langle\phi_{k}|\partial_{\lambda}\phi_{k}\bigr\rangle d\lambda$
is the total dynamical phase. In the limit $i\rightarrow\infty$,
the geometric phase becomes \cite{Mukunda1993}
\begin{align}
\theta_{geo} & =\textrm{Im\,ln}\langle\hat{\mathcal{O}}_{\lambda}\rangle-\underset{i\rightarrow\infty}{\textrm{Lim}}\bigl\{\textrm{Im}\sum_{k=1}^{i-1}\bigl\langle\phi_{k}|\partial_{\lambda}\phi_{k}\bigr\rangle d\lambda\bigr\}\nonumber \\
 & =\textrm{Im\,ln}\langle\hat{\mathcal{O}}_{\lambda}\rangle-\textrm{Im}\varint_{\lambda_{1}}^{\lambda_{i}}\left\langle \!\right.\hat{\mathcal{O}}_{\lambda}^{\dagger}\partial_{\lambda}\hat{\mathcal{O}}_{\lambda}\left.\!\right\rangle d\lambda\nonumber \\
 & =\theta_{tot}-\theta_{dyn},
\end{align}
where the integral is taken along the evolution path $\mathcal{C}$
in projective Hilbert space $P(\mathcal{H})$, the space of density
matrices $\varrho$. This generalization  extends the notion of the Pancharatnam phase beyond cyclic trajectories.
When the parallel transport condition $\left\langle \!\right.\hat{\mathcal{O}}_{\lambda}^{\dagger}\partial_{\lambda}\hat{\mathcal{O}}_{\lambda}\left.\!\right\rangle \rightarrow0$
is achieved (see Appendix \ref{sec:Quantum-parallel-transport}),
the noncyclic Pancharatnam phase $\theta_{tot}$ becomes purely geometric.
Figure \ref{fig:2}(a) shows the postselection probability $\mathcal{P}^{(j)}$
as a function of $\Theta$ for the case of $\hat{\mathcal{O}}_{\lambda}^{P}$
for different $j$ values. All curves are uniformly phase-shifted (by the Pancharatnam phase $\textrm{Im\,ln}\langle\hat{\mathcal{O}}_{\lambda}\rangle$),
resulting in coincident fringe positions that are  independent of $j$.

As a final remark, the noncyclic Pancharatnam phase represents the
most general form in which geometric phases can be defined. Consequently,
our formalism is directly extendable to other scenarios by suitably
modifying the channel operator to incorporate the parameter-dependent
evolution. For example, our framework naturally reduces to the Berry
phase in the special case of adiabatic cyclic evolution \citep{Berry1987}. Specifically,
defining the Berry phase requires specifying a parameter space $\lambda=\lambda(\boldsymbol{R})$
associated with the system, followed by a cyclic evolution of $\lambda$
within this space \footnote{The Berry phase has recently been generalized by the two-time (pre- and
postselected) formalism, making the phase a tunable quantity controlled
by the time separation between the forward-evolving (preselected)
and backward-evolving (postselected) states \citep{Ho2025}.}. In general, we expect that optimization remains effective provided
the Pancharatnam phase shift in the interference pattern is nonzero.
However, if the geometric phase contribution exactly cancels the dynamical
phase, resulting in a net zero phase shift, optimization is expected
to fail.

\end{document}